\documentclass[aps,prl,twocolumn,groupedaddress]{revtex4}
\usepackage{graphicx,color}
\usepackage{amssymb}   
\usepackage{amsmath}
\newcommand{\Mod}[1]{\ (\mathrm{mod}\ #1)}
\usepackage{epstopdf}
\usepackage{natbib}
\usepackage{tabularx,booktabs}
\usepackage[colorlinks,citecolor=green,urlcolor=blue,bookmarks=false,hypertexnames=true]{hyperref} 
\usepackage{bm}
\usepackage{color}
\usepackage{braket}

\begin{document}
\title{Quantum-Geometric Origin of Out-of-plane Stacking Ferroelectricity}
\author{Benjamin T. Zhou} \thanks{Corresponding author: benjamin.zhou@ubc.ca \\ These authors contributed equally to this work.} 
\author{Vedangi Pathak} \thanks{These authors contributed equally to this work.} 
\author{Marcel Franz} 

\affiliation{Department of Physics and Astronomy \& Stewart Blusson Quantum Matter Institute,
University of British Columbia, Vancouver BC, Canada V6T 1Z4}

\begin{abstract}
Stacking ferroelectricity (SFE) has been discovered in a wide range of van der Waals materials and holds promise for applications, including photovoltaics and high-density memory devices. We show that the microscopic origin of out-of-plane stacking ferroelectric polarization can be generally understood as a consequence of nontrivial Berry phase borne out of an effective Su-Schrieffer-Heeger model description with broken sublattice symmetry,
thus elucidating the quantum-geometric origin of polarization in the extremely non-periodic bilayer limit. Our theory applies to known stacking ferroelectrics such as bilayer transition-metal dichalcogenides in 3R and T$_{\rm d}$ phases, as well as general AB-stacked honeycomb bilayers with staggered sublattice potential. Our explanatory and self-consistent framework based on the quantum-geometric perspective establishes quantitative understanding of out-of-plane SFE materials beyond symmetry principles.
\end{abstract}

\maketitle

\emph{Introduction.}--- Two-dimensional (2D) ferroelectrics can serve as building blocks of high-density non-volatile memories~\cite{Garcia, Datta}, but they remain rare among materials found in nature. Recent developments in synthesis of layered van der Waals materials have uncovered a wide range of materials in the bilayer limit~\cite{Wu, Li, DelaBarrera, Yasuda, Stern, Jindal, Xirui, Fei, Yang, Jing, Zheng, Niu, Pacchioni}, called stacking ferroelectrics (SFEs). Intriguingly, the constituent monolayers in SFEs are generally non-polar and spontaneous polarization arises from unusual stacking orders with suitable symmetry breaking conditions allowing the emergence of electric polarity~\cite{Wu, Ji}.

Despite elegant symmetry arguments, a series of important questions regarding the origin of SFE remains unaddressed: according to the modern theory of polarization, the electric polarization stems from the Berry phase encoded in the Bloch wave functions ~\cite{Vanderbuilt, Resta, Resta2, Xiao, Spaldin}. While the in-plane polarization is compatible with a periodic 2D lattice~\cite{Fregoso, CZheng, Higashi}, the extremely non-periodic structure caused by \textit{out-of-plane} polarization $P_z$ in the bilayer limit poses challenges in conceptualizing the polarization as the Berry phase formulated traditionally in momentum-space. Although the magnitude of $P_z$ in the bilayer limit can be obtained by advanced computational methods with non-crystalline generalizations~\cite{RestaPRL, Bennett1, Bennett2}, the Berry phase origin behind the emergence of nonzero $P_z$ has remained non-transparent at the conceptual level. Moreover, existing theoretical studies on SFE have predominantly adopted symmetry-based approach to determine whether $P_z$ exists or not~\cite{Wu, Li, Ji}, whereas symmetry alone does not provide any quantitative understanding of the behavior of $P_z$ in different SFE materials. An explanatory theory which clarifies the Berry phase origin and quantifies the robustness of the out-of-plane SFE is clearly desirable.   

\begin{figure}
    \centering
    \includegraphics[width=0.5\textwidth]{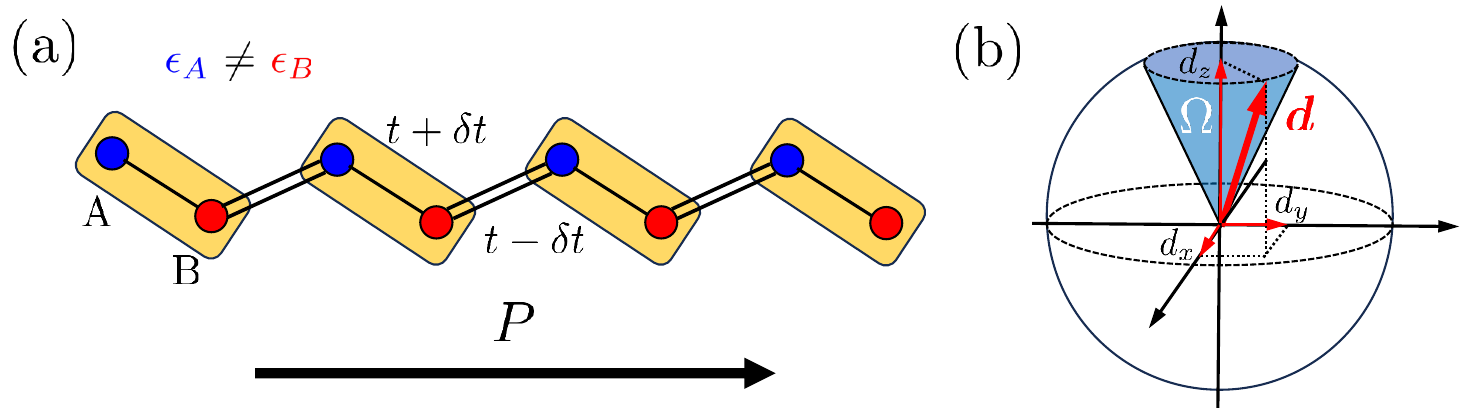}
    \caption{(a) Schematic of a 1D SSH atomic chain with intra-cell hopping $t + \delta t$ and inter-cell hopping $t - \delta t$ between A, B sublattice sites. A polar structure forms under a staggered sublattice potential $\epsilon_A \neq \epsilon_B$. (b) Winding of $\bm{d}$-vector defined in Eq.\ \eqref{eq:SSH} as momentum $k$ is varied adiabatically across the 1D Brillouin zone. Berry phase is given by 1/2 of the solid angle $\Omega$ subtended by $\bm{d}$. }
    \label{FIG1}
\end{figure}

In this Letter, we introduce a mapping between the effective Bloch Hamiltonians of SFEs at each crystal momentum ${\bm p}$ to the \textit{two-cell} limit of the celebrated Su-Schrieffer-Heeger (SSH) chain~\cite{SSH}, characterized under staggered sublattice potentials by a {\em polar} Berry phase. This mapping enables us to develop an explanatory framework which elucidates the Berry phase origin of $P_z$ in various SFE materials and quantifies its robustness through the behavior of the $\bm{d}$-vector characterizing the SSH chain (Fig. \ref{FIG1}). 


\emph{SFE as two-cell limit of sublattice-broken SSH chain.}--- We start by a brief review of the polarization physics in an SSH chain. The SSH model describes a dimerized polyacetylene chain with A/B sublattice sites and alternating bonds (Fig.\ \ref{FIG1}a). The momentum-space Hamiltonian in the Bloch basis $\ket{k, A}, \ket{k, B}$ of the dimerized chain is characterized by a four-component $\bm{d}$-vector
\begin{eqnarray}\label{eq:SSH}
H_{\rm SSH}(k) = \sum_{\alpha = 0,x,y,z} d_{\alpha}(k) \sigma_{\alpha},
\end{eqnarray}
where $d_0(k) = (\epsilon_A + \epsilon_B)/2$, $d_x(k) = (t + \delta t) + (t - \delta t) \cos(ka)$, $d_y(k) = (t - \delta t) \sin(ka)$, $d_z(k) = (\epsilon_A - \epsilon_B)/2$ with $\epsilon_{A}$, $\epsilon_{B}$ as the on-site energies on sublattice A, B, and the Pauli matrices $\sigma_{\alpha}$ act on the sublattice space. According to the modern theory of polarization ~\cite{Vanderbuilt, Resta, Xiao, Spaldin} $P$ of the 1D chain is written as
\begin{equation}\label{eq:BerryPhase}
P = \frac{e}{2\pi} \oint_{k \in {\textrm {BZ}}} \braket{u_{-}(k)|-i\partial_k|u_{-}(k)} dk,   
\end{equation}
where $\ket{u_{-}(k)}$ is the eigenstate of the filled lower band of the two-level Hamiltonian  \eqref{eq:SSH} and the loop integral is precisely the Berry phase $\gamma$ acquired by $\ket{u_{-}(k)}$ across the 1D Brillouin zone (BZ), with $\gamma = \Omega/2$ where $\Omega$ is the solid angle subtended by $\bm{d}$ on the Bloch sphere (Fig.\ \ref{FIG1}b).

\begin{figure*}
    \centering
    \includegraphics[width=\linewidth]{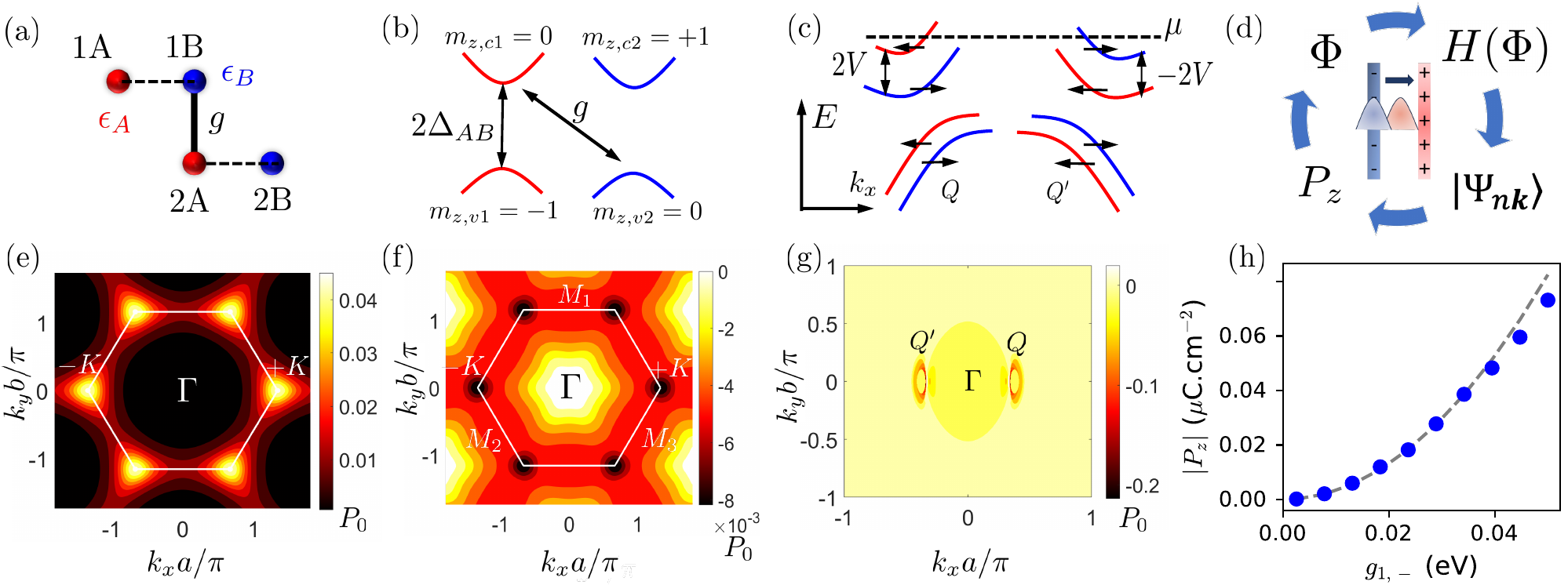}
    \caption{(a) AB-stacked honeycomb bilayer with direct coupling $g$ between 1B and 2B sites. (b) Asymmetric coupling in 3R-bilayer TMD at $+K$ occurs only between the conduction band in layer 1 and the valence band in layer 2 with the same effective angular momentum $m_z = 0$. (c) Schematic band diagram at $Q$- and $Q'$-valleys for bilayer T$_{\rm d}$ TMD. A spin-valley-dependent band splitting $2V$ leads to imbalance between spin subbands from $Q$ and $Q'$ valleys. (d) Feedback loop for the self-consistency condition established in Eq.\ \eqref{eq:SelfConEqn}. (e-g) $\bm{k}$-resolved polarization for (e) bilayer SiC, (f) 3R-bilayer MoS$_2$, (g) T$_{\rm d}$-bilayer WTe$_2$. Color scales in (e,g) represent polarization in units of polarization quantum $e/A_{\rm Cell}$, with $A_{\rm Cell}$ the area of the unit cell. $|\Psi_1|^2$ in (f) denotes weight of eigenstates in layer 1. (h) $P_z$ of bilayer WTe$_2$ as a function of inversion-symmetric $g_{1,-}$.The dashed gray line depicts a quadratic fit of the polarization data.}
    \label{FIG2}
\end{figure*}
In the usual setting $H_{\rm SSH}$ has a sublattice symmetry $\epsilon_A = \epsilon_B$ which implies \textit{global} inversion symmetry: $\mathcal{I} H_{\rm SSH}(-k) \mathcal{I}^{-1} = H_{\rm SSH}(k)$, with the inversion operator $\mathcal{I} = \sigma_x$. The $\mathcal{I}$-symmetry enforces $\gamma = 0, \pi$ (modulo $2\pi n$, $n \in \mathbb{Z}$ due to the ambiguity of Berry phase in 1D), which characterize \textit{non-polar} phases~\cite{Resta, Spaldin}. A simple way to polarize the SSH chain is to introduce a staggered sublattice potential $\Delta_{AB} \equiv (\epsilon_A - \epsilon_B)/2 \neq 0$ such that $\mathcal{I}$ is broken and $d_{\alpha} \neq 0$ for all $\alpha = x,y,z$, which is also known as the Rice-Mele model~\cite{RiceMele}. $\Omega$ now takes on a general value $\Omega \in (0, 2\pi)$ and $\gamma = \Omega/2 \in (0, \pi)$ indicates non-vanishing polarization. 

The relevant parameter choice for SFE corresponds to the  limit where intra-cell bonding vanishes, $t = -\delta t$. The Berry phase then assumes a simple form
\begin{align}\label{eq:BerryPhaseSSH}
\gamma = \pi \left[1 - \frac{\Delta_{AB}}{ \sqrt{4t^2 + \Delta_{AB}^2}}   \right] \Mod{2n\pi, n \in \mathbb{Z}}.
\end{align}
Note that the polarity of the SSH chain is robust when $\gamma$ is kept away from the non-polar values $0$ and $\pi$, which is attained for $\Delta_{AB} \simeq 2t$ according to Eq.\ \eqref{eq:BerryPhaseSSH}. This result can be understood pictorially as follows: the non-polar $\gamma = 0$ phase corresponds to $\Delta_{AB} \gg 2t$ where the $\bm{d}$-vector is pinned at the poles of the Bloch sphere, while the non-polar $\gamma = \pi$ phase is attained for $\Delta_{AB} \ll 2t$ where $\bm{d}$ stays on the equator. With $\Delta_{AB} \simeq 2t$, the $\bm{d}$-vector lies midway between the poles and the equator, and the system is firmly embedded in the polar phase. Next, we discuss how the polarization in SFE materials can be interpreted as a consequence of the Berry phase in Eq.\ \eqref{eq:BerryPhaseSSH}.

(i) \emph{Type-I SFEs: AB-stacked honeycomb bilayers.}---A prototypical class of SFE materials is the AB-stacked bilayers with honeycomb lattice symmetry, such as hexagonal boron nitride (hBN), gallium nitride (GaN), and silicon carbide (SiC)~\cite{Wu, Yasuda, Stern}. The crystalline structure of each constituent monolayer has the non-polar $D_{3h}$-point group, which resembles graphene with intrinsically broken AB sublattice symmetry. The AB stacking breaks the horizontal mirror symmetry $\sigma_{h}$, which further reduces the symmetry to the polar $C_{3v}$ point group compatible with a nonzero $P_z$. 

To elucidate the connection between Eq.\ \eqref{eq:BerryPhaseSSH} and $P_z$, we note that the low-energy effective bilayer Hamiltonian in the Bloch basis of $p_z$-orbitals on A1, B1, A2, B2 sites takes the general form
\begin{eqnarray}\label{eq:ABBilayer}
H_{AB,\xi}(\bm{p}) = 
\begin{pmatrix}
\epsilon_{A1} & v p_{\xi,-}   & v_1 p_{\xi,-}   & v_2 p_{\xi,+} \\
v p_{\xi,+} & \epsilon_{B1}   & g           & v_1 p_{\xi,-} \\
v_1 p_{\xi,+} & g             & \epsilon_{A2} & v p_{\xi,-} \\
v_2 p_{\xi,-}  & v_1 p_{\xi,+}     & v p_{\xi,+}     & \epsilon_{B2}
\end{pmatrix},
\end{eqnarray}
where $p_{\xi,\pm} = \xi p_{x} \pm i p_y$ is measured from the two inequivalent $K$-points indexed by $\xi = \pm$. $\epsilon_{ml}$ is the on-site energy on sublattice $m=\textrm{A, B}$ and layer $l=1,2$ with $\epsilon_{A,l} \neq \epsilon_{B,l}$ due to different atoms on AB sublattices, $g$ is the direct inter-layer hopping between B1 and A2 sites (Fig.\ \ref{FIG2}a). By mapping the AB sublattice in each layer to the AB sublattice in each cell of the SSH chain, Eq.\ \eqref{eq:ABBilayer} near the $K$-points with $\bm{p} \simeq \bm{0}$ becomes simply the SSH chain with two unit cells (Fig.\ \ref{FIG1}), where the intra-cell hopping $ t + \delta t \equiv v p_{\xi,-} \simeq 0$, and the inter-cell hopping between adjacent AB sites $t - \delta t = 2t \equiv g$. By extending the bilayer Hamiltonian to the 3D limit with the number of layers $N_z \rightarrow \infty$, $P_z$ is precisely the integral over $k_z$ in Eq.\ \eqref{eq:BerryPhase} and $g$ enters Eq.\ \eqref{eq:BerryPhaseSSH} as $2t$. Since the origin of $P_z$ does not depend on $N_z$, $\gamma$ obtained in the large $N_z$ limit necessarily implies a nonzero $P_z$ with the same origin in the bilayer ($N_z = 2$).

(ii) \emph{Type-II SFEs: Rhombohedral (3R) bilayer TMDs.}---The 3R-structure of bilayer TMDs is formed by two 2H-monolayers assembled with the rhombohedral stacking order~\cite{Maschmeyer}. Despite a similar crystalline structure as AB-stacked honeycomb bilayers, the 3R-bilayer TMDs have different basis states than the $p_z$-orbitals: the conduction band states $\ket{c,\pm K,l}$ and valence band states $\ket{v,\pm K,l}$ at $K, K'$ originate from transition-metal $d$-orbitals with different angular momenta $m_z$~\cite{GuiBin, Yao, Kormanyos} (Fig.\ \ref{FIG2}b). The AB stacking order causes a relative shift in $m_z$ at $\pm K$ between states from different layers such that $m_z$ for states in different layers have an extra difference of $\pm 1$~\cite{Yao2, Yang, Jing}. As such, the $\mathcal{C}_{3z}$ symmetry enforces the inter-layer coupling at $\pm K$ to be \textit{asymmetric} as tunneling is allowed only between $\ket{c,\pm K,1}$ and $\ket{v,\pm K,2}$ with the same $\mathcal{C}_{3z}$ eigenvalues (Fig.\ \ref{FIG2}b). By mapping the conduction (valence) band states in each layer to the A (B) sublattice in each cell of the SSH chain, the effective Hamiltonian near $\pm K$ has a form similar to Eq.\ \eqref{eq:ABBilayer}, where $\Delta_{AB}\equiv (E_{c} - E_{v})/2$ is half of the direct semiconducting gap - see detailed derivation in the Supplemental Material (SM)~\cite{SM}.  The system is thus characterized by a polar Berry phase of the form Eq.\ \eqref{eq:BerryPhaseSSH} following similar analysis in subsection (i). 

(iii) \emph{Type-III SFEs: Bilayer $\rm{T}_{\rm d}$-structure TMDs.}---A bilayer T$_{\rm d}$-structure TMD is formed by two centrosymmetric topological 1T'-monolayers stacked with a relative $\mathcal{C}_{2z}$ rotation~\cite{Car, Qiong}. Due to its extremely low $C_{s}$ point group symmetry, the T$_{\rm d}$-bilayer exhibits out-of-plane polarity and was among the first sliding ferroelectrics discovered~\cite{Fei}. The low-energy physics in both 1T'-monolayer and T$_{\rm d}$-bilayer involves states near the $Q$ and $Q'$ points of the BZ where the inter-orbital spin-orbit coupling (SOC) opens up nontrivial band gaps at the topological band crossing points~\cite{Qiong, Du, Junwei, Sanfeng, Benjamin}. Using a symmetry-adapted $\bm{k}\cdot\bm{p}$ model for T$_{\rm d}$-bilayer~\cite{SM, Junwei, Benjamin}, we find that the bilayer near $Q, Q'$ can be described by asymmetrically coupled massive Dirac fermions similar to Eq.\ \eqref{eq:ABBilayer}, with spin-valley-dependent Dirac masses $m_{\xi\sigma} = \xi \sigma m$ generated by the inter-orbital SOC (details in SM~\cite{SM}). The problem then can be mapped to the two-cell limit of a valley-dependent \textit{spinful} SSH chain, which in the $N_z \rightarrow\infty$ limit is characterized by a $\bm{d}$-vector in Eq.\ \eqref{eq:SSH} with components $d_{0, \xi\sigma} = \xi\sigma V$, $d_x(k) = 2g_{0,-}\cos(ka), d_y(k) = -2g_{1,-}\sin(ka)$, $d_{z, \xi\sigma}(k) = \xi\sigma m + 2 g_{1,+} \cos(ka)$. Here, the spin-valley-dependent potential $\xi\sigma V$ arises from the combination of SOC and broken $\mathcal{I}$-symmetry in T$_{\rm d}$ bilayer and lifts the spin degeneracy (Fig.\ \ref{FIG2}c). The $g_{0,\pm}, g_{1,-}$ terms are even under $\mathcal{I}$ while the $g_{1,+}$ term is odd~\cite{SM}.

For illustration we consider the simple limit $|g_{0,-}| = |g_{1,-}|\equiv g$ which enables analytic calculation of the spin-dependent Berry phase and gives, for $m\gg g$ under realistic settings (see SM~\cite{SM} for derivation):
\begin{equation}\label{eq:SpinBerryPhase}
\gamma_{\xi\sigma} \simeq  \xi\sigma \frac{m g^2 \pi}{(m^2 - 4g_{1,+}^2)^{3/2}} . 
\end{equation}
Note that $\gamma_{\xi, +} \simeq -\gamma_{\xi, -}$ at each valley $\xi$ due to the spin-dependent sign in $m_{\xi\sigma}$, while for general band filling the different spin subbands in the conduction band states have different occupation numbers due to the finite spin-splitting $V$ (Fig.\ \ref{FIG2}c), which results in a partial cancellation between different spin sectors only and the net Berry phase remains finite (see SM~\cite{SM} for details).

Notably, the polar Berry phases in Eq.~\eqref{eq:BerryPhaseSSH} and Eq.~\eqref{eq:SpinBerryPhase} obtained in the $N_z \rightarrow \infty$ limit accurately account for the bare polarization $P_{z,0}$ near $K$ and $Q$ points generated by stacking geometry in the bilayer limit (see Section II of SM~\cite{SM}). Moreover, by constructing realistic lattice models for bilayer SFEs that incorporate Bloch states at all in-plane $\bm{k}$, we find that the bare layer polarization $P_{z,0}$ is indeed dominated by contributions near $K$ and $Q$ points (see Section II-2C and Table S6 of SM~\cite{SM}). This reveals that the Berry phase given by Eq.~\eqref{eq:BerryPhaseSSH} and Eq.~\eqref{eq:SpinBerryPhase} above provides the primary source of the out-of-plane polarization in the bilayer SFEs discussed above.

\setlength{\tabcolsep}{6pt} 
\begin{table*}[t]
\centering 
\caption{Comparison among different types of SFE materials. Specific examples of type I-III SFEs are represented by bilayer SiC, bilayer 3R-MoS$_2$ and bilayer T$_{\rm d}$-WTe$_2$, respectively. Reported values of $P_z$ (last row) are taken from previous DFT studies~\cite{Wu,Li} and experiments~\cite{Yang, DelaBarrera, Fei}. See Supplemental Material~\cite{SM} for details of model parameters.}
\begin{tabular}{c c c c}
\hline\hline
Type of SFE & AB-stacked honeycomb bilayer & 3R-bilayer TMD & T$_{\rm d}$-bilayer TMD\\
\hline
Examples &            BN, GaN, SiC         &  MoS$_2$, MoSe$_2$, WS$_2$ &  WTe$_2$, MoTe$_2$\\
\hline
$\Delta_{AB}$ in SSH chain & Staggered sublattice potential & Semiconducting gap  & Inter-orbital SOC \\
\hline
Asymmetric coupling &  Direct $\sigma$-bond (1B - 2A) &  Tunneling between $\ket{v,1}$, $\ket{c,2}$  &  Inter-orbital tunneling\\
\hline
Robustness of polarity & Strong   &    Intermediate    &      Weak\\
\hline
$P_z$ (obtained from Eq.\ \eqref{eq:SelfConEqn}) &  1.77 $\rm \mu C /{\rm cm}^{2}$ (SiC) &  0.6 $\rm \mu C /{\rm cm}^{2}$ (MoS$_2$)  &  0.034 $\rm \mu C/ {\rm cm}^{2}$ (WTe$_2$) \\
\hline
$P_z$ (reported in literature) &  1.76 $\rm \mu C /{\rm cm}^{2}$ (SiC) &  0.6 $\rm \mu C/{\rm cm}^{2}$ (MoS$_2$)  &  0.02-0.06 $\rm \mu C/{\rm cm}^{2}$ (WTe$_2$) \\
\hline\hline
\end{tabular}
\label{table:table1}
\end{table*}

\emph{Self-consistent formalism with surface charge corrections}. --- It is worth noting that the actual value of $P_z$ for a non-periodic bilayer SFE is not determined by the bare $P_{z,0}$ generated by stacking geometry alone, but is subject to corrections from surface bound charge effects. In usual Berry phase calculations for bulk ferroelectrics, surface charges can be safely neglected as they are well separated in space from the interior of the bulk, which justifies the calculation of bulk $P_z$ based on Berry connections defined for a periodic system with no physical boundaries and a vanishing macroscopic electric field~\cite{Resta2}. In the bilayer limit, however, surface effects can no longer be ignored --- charges of opposite signs are located right on the top and bottom layers that constitute the bilayer system. This necessarily induces a non-negligible inter-layer potential difference $\Phi \neq 0$ in the Hamiltonian, which reduces the bare polarization $P_{z,0}$ and leads to a corrected value of $P_z$ through the feedback loop depicted schematically in Fig.~\ref{FIG2}d.

For $P_z$ and $\Phi$ to be well-defined physical quantities, the value of $P_z$ obtained from the microscopic quantum theory under a given $\Phi \neq 0$ should generate the same value $\Phi = -e P_z d/\epsilon$ of the potential  required by electrostatics, where $\epsilon$ is the dielectric constant and $d$ the inter-layer spacing. Thus, $P_z$ must be determined self-consistently via the equation (Section II-3 in SM~\cite{SM}):
\begin{eqnarray}\label{eq:SelfConEqn}
 P_z = -\frac{e}{\mathcal{V}} \sum_{n, \bm{k} \in {\rm BZ}}  \braket{\Psi_{n\bm{k}}(\Phi)|\hat{z}|\Psi_{n\bm{k}}(\Phi)}, 
\end{eqnarray}
 where $\ket{\Psi_{n \bm{k}}(\Phi)}$ denotes the Bloch state of a filled band $n$ at in-plane momentum $\bm{k}$ under a given $\Phi$. For concreteness, we consider bilayer SiC, 3R bilayer MoS$_2$ and T$_{\rm d}$-bilayer WTe$_2$ as specific examples for the three types of SFE and present our results in Table~\ref{table:table1}. Excellent agreement is found between the results obtained by self-consitently solving Eq.\ \eqref{eq:SelfConEqn} and literature values. 
 
It is important to emphasize that while $P_{z} \neq P_{z,0}$ in general, the primary source of the out-of-plane polarity must always be attributed to the bare $P_{z,0}$ generated by nontrivial stacking geometry, which arises primarily from the polar Berry phases in Eq.~\eqref{eq:BerryPhaseSSH} and Eq.~\eqref{eq:SpinBerryPhase}. Corrections from $\Phi\neq 0$ should be viewed as a secondary effect which would not have existed if $P_{z,0} =0$ (or equivalently, $\gamma = 0, \pi$) in the first place under a non-polar stacking geometry. Importantly, corrections from $\Phi \neq 0$ do not affect the dominant role of contributions from $K$ and $Q$ points as shown in the $\bm{k}$-resolved polarization in Fig.\ \ref{FIG2}e-g: for $\Phi\neq 0$, contributions to total $P_z$ remain concentrated in the neighborhood of the $K$-points in AB-stacked honeycomb bilayers and 3R-bilayer TMDs, and near the $Q$-points in T$_{\rm d}$-TMDs.

For 3R-bilayer MoS$_2$ some minor but non-negligible contributions to $P_z$ do arise from regions around the time-reversal-invariant $M_1, M_2, M_3$ points (Fig.~\ref{FIG2}f). As we discuss in detail in Section III of the SM~\cite{SM}, the effective Hamiltonian near these $M$-points can also be mapped to a two-cell SSH chain. The nonzero $P_{z,0}(\bm{k})$ originates from polar Berry phases of the same form as Eq.~\eqref{eq:BerryPhaseSSH} which reinforces our claim that the polarization in 3R-MoS$_2$ is rooted in the SSH physics.

 

\emph{Comparison among SFE materials.}--- Following the criteria we established via Eq.\ \eqref{eq:BerryPhaseSSH}, the electric polarity is robust when $\Delta_{AB} \sim g \equiv 2t$. In the context of SFEs, parameter $\Delta_{AB}$ is usually given by the intrinsic band gap and $g$ is the asymmetric inter-layer coupling strength (Eq. \ref{eq:ABBilayer}). In the specific case of SiC we consider for AB honeycomb bilayers of type (I), the intrinsic band gap is of order 2 eVs, and $g$ originates from the strong $\sigma$-bond between $p_z$-orbitals of order 0.5 eV~\cite{SM, McCann}. Thus, SiC is within the $\Delta_{AB} \sim g$ regime and we expect similar physics in other AB-stacked bilayers with strong $\sigma$-bonds. For 3R-bilayer TMDs of type (II), the inter-layer hopping is relatively weaker $g \sim 0.1$ eV~\cite{Yang, Jing, SM} which places the polarity of these materials in the intermediate range. 

The polarity of T$_d$-bilayer TMD is the weakest among the three types due to the partial cancellation between $\gamma_{\xi, +}, \gamma_{\xi, -}$, and the fact that the emergence of $P_z$ relies on nonzero $V$ due to broken inversion. On the other hand, $\mathcal{I}$-symmetry breaking is a necessary but insufficient condition for electric polarity, and the role of geometry is essential: if $\gamma_{\xi\sigma}$ in Eq.\ \eqref{eq:SpinBerryPhase} is zero, the system should exhibit no polarity even if $\mathcal{I}$ is broken. We demonstrate this by artificially tuning the strength of the $\mathcal{I}$-preserving $g_{1,-}$-term, which can modify the Berry phase according to Eq.\ \eqref{eq:SpinBerryPhase} but keeps the symmetry of the system unchanged. As shown clearly in Fig.\ \ref{FIG2}h, $P_z$ is negligible for $g_{1,-} = 0$ where $\mathcal{I}$ is already broken, while increases monotonically as a function of $g_{1,-}$. This exemplifies the essential role of Berry phase for understanding SFE.
 
\emph{Conclusion and Discussions}.--- Our considerations uncover the quantum-geometric origin of SFE polarization by establishing its close relation to the geometric property of the $\bm{d}$-vector characterizing the classic SSH chain. This geometric approach not only clarifies the Berry phase origin behind out-of-plane SFE, but also establishes a general criterion, \textit{i.e.}, $\Delta_{AB} \sim g$, as the key condition for robust out-of-plane SFE polarization, which goes beyond the current symmetry-based understanding of SFE materials. 

By identifying the origin of $P_z$ from Berry phase of the form in Eqs.\ \eqref{eq:BerryPhaseSSH} and \eqref{eq:SpinBerryPhase}, our theory reveals an important \textit{correlation} among the key parameters characterizing bilayer SFE: the inter-layer potential difference $\Phi$ must be correlated with parameters $g$ and $\Delta_{AB}$ which generate the Berry phase behind the nonzero $P_z$. This stands in contrast to existing effective models for the bilayer out-of-plane SFE where a finite $\Phi$ is introduced in 3R-bilayer TMD but the asymmetric inter-layer coupling $g$ is omitted~\cite{Burkard}, or a symmetric coupling form is assumed in a toy model for T$_{\rm d}$-bilayer WTe$_2$~\cite{Du}. 

The correlation among the parameters discussed above has further implications when there is an interplay between electron interactions and out-of-plane ferroelectricity -- in a recent experiment the superconducting state in T$_{\rm d}$-bilayer MoTe$_2$ is found to exhibit out-of-plane ferroelectric hysteresis, which suggests strong coupling between the superconducting order and the out-of-plane ferroelectric order~\cite{Rhodes}. Since superconducting pairing in T$_{\rm d}$ TMDs occurs between electrons from the $Q, Q'$-valleys~\cite{Benjamin, Crepel} where the Berry phase is most prominent, it follows that the superconducting order parameters must also be determined self-consistently with respect to parameters entering the Berry phase in Eq.\ \eqref{eq:SpinBerryPhase}. Our self-consistent formalism thus provides a minimal reliable normal-state description for future studies of the ferroelectric superconductivity in T$_{\rm d}$-bilayer TMDs.

We conclude by noting that our finding of dominant contribution to the Berry phase from the vicinity of the $K$ and $Q$ points (Fig.\ \ref{FIG2}) provides useful guidance for photo-current measurements on bilayer SFE: since the shift current scales with the inter-band polarization difference~\cite{Rappe, Moore, Cook}, the relevant optical transitions to look for in experiments would be those around $K$ and $Q$ where the band edges possess different layer-polarizations (\textit{e.g.}, Fig.\ \ref{FIG2}f), which allows incident photons to pump electrons between the layers.  
\\

\emph{Acknowledgement.} --- We thank Dongyang Yang, Jing Liang and Ziliang Ye for illuminating discussions and fruitful collaborations which inspired the current work. We further thank Hongyi Yu and Daniel Bennett for helpful discussions through email correspondence. This work was supported by NSERC, CIFAR and the Canada First Research Excellence Fund, Quantum Materials and Future Technologies Program. \\

\emph{Note added.} --- After submission of this manuscript, we became aware of a recent work~\cite{Hongyi} on bilayer hBN which reports $\bm{k}$-resolved inter-layer charge redistribution with similar features we found for the $\bm{k}$-resolved polarization in bilayer SiC (Fig. \ref{FIG2}e). 


\clearpage

\onecolumngrid
\begin{center}
\textbf{\large Supplemental Material for\\ ``Quantum Geometric Origin of Out-of-plane Stacking Ferroelectricity''} \\[.2cm]
Benjamin T. Zhou,$^{1}$ Vedangi Pathak,$^{1}$ Marcel Franz$^{1}$\\[.1cm]
{\itshape ${}^1$Department of Physics and Astronomy \& Stewart Blusson Quantum Matter Institute,
University of British Columbia, Vancouver BC, Canada V6T 1Z4}	\\
\end{center}
\setcounter{equation}{0}
\setcounter{section}{0}
\setcounter{figure}{0}
\setcounter{table}{0}
\setcounter{page}{1}
\renewcommand{\theequation}{S\arabic{equation}}
\renewcommand{\thetable}{S\arabic{table}}
\renewcommand{\thesection}{\arabic{section}}
\renewcommand{\thefigure}{S\arabic{figure}}
\renewcommand{\bibnumfmt}[1]{[S#1]}
\renewcommand{\citenumfont}[1]{S#1}
\makeatletter

\section*{I. Microscopic models of SFE materials}

In this section, we present details of the microscopic models for the three types of SFE materials discussed in the main text. In particular, we demonstrate how the effective models near $K$ and $Q$, which capture the SSH physics in SFE materials, are derived from realistic tight-binding models. We take bilayer SiC, bilayer 3R-MoS$_2$ and T$_{\rm d}$-bilayer WTe$_2$ as specific examples below for illustration, while the analysis applies in general to other materials belonging to each type of SFE. For simplicity, we ignore the the inter-layer potential difference $\Phi$ within this section, while details of the self-consistent calculation of polarization and $\Phi$ are discussed in Section II of this Supplemental Material.

\subsection*{1. AB-stacked honeycomb bilayers}

\subsubsection*{A. Tight-binding model}

A monolayer SiC can be regarded as a monolayer graphene with broken sublattice symmetry due to different Si and C atoms on AB sublattice sites. The Bravais lattice of SiC is a triangular lattice with lattice vectors $\mathbf{a_1}=a(1,0)$ and $\mathbf{a_2}=a(\frac{1}{2}, \frac{\sqrt{3}}{2})$ where $a$ is the lattice constant. The sublattice A has three nearest neighbour sublattice B sites, which are connected through the bonding vectors $\bm{\delta_1}=a\left(0,\frac{1}{\sqrt{3}}\right)$, $\bm{\delta_2}=a \left(\frac{1}{2},-\frac{1}{2\sqrt{3}}\right)$ and $\bm{\delta_3}=a\left(-\frac{1}{2},-\frac{1}{2\sqrt{3}}\right)$. The tight-binding Hamiltonian for a monolayer SiC can be written as
\begin{equation}
    H=-t\sum_{\langle ij\rangle} \left(c_{iA}^{\dag}c_{jB}+h.c.\right)+\epsilon_A\sum_ic_{iA}^{\dag}c_{iA}+\epsilon_B\sum_ic_{iB}^{\dag}c_{iB},
\end{equation}
where $t$ is the nearest neighbour hopping term and $\Delta_{A(B)}$ is the onsite potential of atoms in sublattice A (B). The momentum-space Hamiltonian in the Bloch basis $c^{\dag}_{\bm{k}}=\left(c^{\dag}_{A\bm{k}},c^{\dag}_{B\bm{k}}\right)$ is given by
\begin{align}
H&=\sum_{\bm{k}} c^{\dag}_{\bm{k}}h(\bm{k})c_{\bm{k}}, \ \ \ \ \
 h(\bm{k})=
\begin{pmatrix}
\epsilon_A & -t f(\bm{k})\\
-t f^*(\bm{k}) &\epsilon_B
\end{pmatrix}, 
\end{align}
where $f(\bm{k})=\sum_{j=1,2,3}\text{e}^{-i\bm{k}\cdot\bm{\delta_j}}$. For an AB-stacked bilayer, we have sublattice $A_1, B_1$ and $A_2, B_2$ on the top and bottom layers respectively. For the $A_2 - B_1$ stacking configuration, the interlayer coupling matrix can be written as
\begin{align}
    T(\bm{k})&=
    \begin{pmatrix}
        T_{A_1, A_2}&T_{A_1, B_2}\\T_{B_1, A_2}&T_{B_1, B_2}
    \end{pmatrix},
\end{align}
where $T_{A_1, A_2} =-g_{AA} f(\bm{k})$, $T_{A_1, B_2} =-g_{AB} f^*(\bm{k})$, $T_{B_1, A_2} =g_{BA}$, $T_{B_1, B_2} =-g_{BB} f(\bm{k})$. In the basis $c^{\dag}_{\bm{k}}=\left(c^{\dag}_{A'\bm{k}},c^{\dag}_{B'\bm{k}},c^{\dag}_{A\bm{k}},c^{\dag}_{B\bm{k}}\right)$ the total bilayer Hamiltonian matrix is given by
\begin{align}\label{eq:tb_sic}
H_{AB}(\bm{k})&=
\begin{pmatrix}
h(\bm{k})&T(\bm{k})\\
T^{\dag}(\bm{k})&h(\bm{k})
\end{pmatrix}.
\end{align}

\begin{table}[]
\caption{\label{tab:sic1} Tight-binding parameters in Eq.~\eqref{eq:tb_sic} for AB-stacked bilayer SiC. The parameters for inter-layer coupling are obtained by fitting  band structures from first-principle calculations for bilayer SiC~\cite{sic_dftS}. Units: eV. }
\centering
\begin{tabular}{lllllll}\hline\hline
$t$   & $g_{AA}$ & $g_{AB}$ & $g_{BA}$ & $g_{BB}$ & $\epsilon_A$ & $\epsilon_B$ \\\hline
1.50 & 0.084       & 0.084       & 0.4       & 0.084       & 1.50                        & -1.00     \\\hline\hline                   
\end{tabular}
\end{table} 

\subsubsection*{B. Effective SSH model at $\pm K$}

It is straightforward to show that the expansion of $H_{AB}(\mathbf{k})$ in the neighborhood of $K$-points, with $\bm{p} = \mathbf{k} - \xi \bm{K}$ ($\xi = \pm$: valley index) measured from the $\xi \bm{K}$ points, leads to the effective Hamiltonian in Eq.\ (4) of the main text. In particular, right at the $\pm \bm{K}$-point we have $f(\bm{k} = \pm\bm{K})=0$, and the momentum-space Hamiltonian becomes
\begin{eqnarray}\label{eq:tb_sic_k}
H_{AB}(\bm{K}) = 
\begin{pmatrix}
\epsilon_{A} & 0   & 0  & 0 \\
0 & \epsilon_{B}   & g_{BA}           & 0\\
0 & g_{BA}             & \epsilon_{A} & 0 \\
0  & 0     & 0    & \epsilon_{B}
\end{pmatrix}.
\end{eqnarray}
As we discussed in the main text, this simply describes the two-cell limit of the classic Su-Schrieffer-Heeger (SSH) chain \cite{SSHS}, with a sublattice asymmetry $\epsilon_A \neq \epsilon_B$ in the ``dimer" in each unit cell. By stacking a large number of SiC monolayers with AB configuration along the $z$-direction, the corresponding effective bulk Hamiltonian in the $N_z \rightarrow \infty$ limit at $\pm \bm{K}$ is given by
\begin{eqnarray}\label{eq:heff_berry}
H_{\pm K} (k_z) =
\begin{pmatrix}
\epsilon_{A} & g_{BA} e^{- i k_z c}\\
g_{BA} e^{ i k_z c} & \epsilon_{B}
\end{pmatrix}
=\frac{\epsilon_A + \epsilon_B}{2} \tau_0 + \frac{\epsilon_A - \epsilon_B}{2} \tau_z + g_{BA} \cos(k_z c) \tau_x +  g_{BA} \sin(k_z c) \tau_y,
\end{eqnarray}
where the Pauli matrices $\tau_{i = 0,x,y,z}$ act on the sublattice space, and $c$ is the lattice spacing along the vertical $c$-axis. The $\bm{d}$-vector characterizing the two-level system is 
\begin{equation}\label{eq:d_sic}
\bm{d}(k_z) = (g_{BA} \cos(k_z c), g_{BA} \sin(k_z c),\frac{\epsilon_A - \epsilon_B}{2}),    
\end{equation}
and the Berry phase of $H_{\pm K} (k_z)$ in Eq.~\eqref{eq:tb_sic_k} is given by: 
\begin{equation}\label{eq:berry_sic}
\gamma = \pi \left(1-\frac{\epsilon_A - \epsilon_B}{2\sqrt{g_{BA}^2 + \frac{(\epsilon_A - \epsilon_B)^2}{4}}}\right).    
\end{equation}
By identifying $\Delta_{AB} \equiv \frac{\epsilon_{A} - \epsilon_B}{2}$ and $2t = g_{BA}$ as the staggered sublattice potential and inter-cell bonding in the SSH chain with broken intra-cell bonding, Eq.~\eqref{eq:berry_sic} reproduces the Berry phase in Eq.3 of the main text.

\subsection*{2. Rhombohedral (3R) bilayer TMDs}

Here, we present a realistic 12-band tight-binding model for bilayer 3R MoS$_2$ constructed based on a third-nearest-neighbor (TNN) tight-binding model for monolayer MoS$_2$ \cite{GuiBinS}. The inter-layer coupling terms are derived using the $C_{3v}$ point group symmetries of 3R MoS$_2$. We explicitly demonstrate that in the neighborhood of $\pm K$, the realistic tight-binding model reduces to an effective Hamiltonian similar to Eq.\ (4) of the main text for AB-stacked honeycomb bilayers.

\subsubsection*{A. Tight-binding model}

According to Ref.\ \cite{GuiBinS}, states in the lowest conduction and topmost valence bands in monolayer MoS$_2$ are dominated by $d_{z^2}, d_{xy}, d_{x^2-y^2}$ orbitals from the triangularly arranged molybdenum atoms. To model bilayer 3R MoS$_2$, we first consider two decoupled monolayers separated by inter-layer spacing $d_z$ and choose the basis states to be the set of Bloch states formed by localized orbitals of $d_{z^2}, d_{xy}, d_{x^2-y^2}$ characters in each layer. The basis Bloch wave functions have the general form 
\begin{eqnarray}
\Psi^{(l)}_{\sigma \alpha \bm{k} }(\bm{r}, z) = \frac{1}{\sqrt{N}}  \sum_{\bm{R}^{(l)}} e^{i \bm{k}\cdot\bm{R}^{(l)}} \phi_{\sigma\alpha} (\bm{r} - \bm{R}^{(l)}, z - R^{(l)}_z),
\end{eqnarray} 
where $\sigma = \uparrow, \downarrow$ is the spin index, $\alpha = d_{z^2}, d_{xy}, d_{x^2-y^2}$ is the orbital index, and $l = 1,2$ is the layer index. $\bm{r} = (x,y)$ labels the 2D spatial coordinate, $z$ labels the coordinate along the vertical $c$-axis, $N$ refers to the total number of sites in the 2D triangular lattice. $\phi_{\sigma\alpha} (\bm{r} - \bm{R}^{(l)}, z - R^{(l)}_z)$ denotes a localized orbital with spin $\sigma$ and orbital character $\alpha$ located at site $(\bm{R}^{(l)}, R_z^{(l)})$, where $\bm{R}^{(l)}$ labels the 2D triangular lattice sites for layer $l$ while $R_z^{(l)} = (-1)^{(l-1)} \frac{d_z}{2}$ labels the sites along the $c$-axis. In the basis of $\{ \ket{ \Psi^{(1)}_{\uparrow, d_{z^2}, \bm{k} } }, \ket{ \Psi^{(1)}_{\uparrow, d_{xy}, \bm{k} }}, \ket{ \Psi^{(1)}_{\uparrow, d_{x^2-y^2}, \bm{k} }}, \ket{ \Psi^{(1)}_{\downarrow, d_{z^2}, \bm{k} }}, \ket{\Psi^{(1)}_{\downarrow, d_{xy}, \bm{k} }}, \ket{ \Psi^{(1)}_{\downarrow, d_{x^2-y^2}, \bm{k} }}$, $\ket{ \Psi^{(2)}_{\uparrow, d_{z^2}, \bm{k} } }, \ket{ \Psi^{(2)}_{\uparrow, d_{xy}, \bm{k} }}, \ket{ \Psi^{(2)}_{\uparrow, d_{x^2-y^2}, \bm{k} }}, \ket{ \Psi^{(2)}_{\downarrow, d_{z^2}, \bm{k} }}, \ket{\Psi^{(2)}_{\downarrow, d_{xy}, \bm{k} }}, \ket{ \Psi^{(2)}_{\downarrow, d_{x^2-y^2}, \bm{k} }} \}$, the 12-band TB Hamiltonian for bilayer 3R MoS$_2$ is written as
\begin{eqnarray}\label{eq:H3R}
H_{3R}(\bm{k}) = 
\begin{pmatrix}
H_{0}(\bm{k}) & T_{R}(\bm{k})\\
T^{\dagger}_{R}(\bm{k}) & H_{0}(\bm{k}) 
\end{pmatrix}.
\end{eqnarray}
Here, $H_0(\bm{k})$ denotes the third-nearest-neighbor(TNN) tight-binding Hamiltonian for monolayer MoS$_2$ \cite{GuiBinS}, and $T_{R}(\bm{k})$ denotes the inter-layer coupling terms, where the subscript $R$ refers to rhombohedral stacking. The form of $H_0(\bm{k})$ is explicitly given by
\begin{eqnarray}
H_0(\bm{k})&=& \sigma_0 \otimes H_{\text{TNN}}\left(\bm{k}\right)+\frac{1}{2}\lambda \sigma_z \otimes L_z, \\\nonumber
\end{eqnarray}
where $\sigma_0$ and $\sigma_z$ denote the usual Pauli matrices for spin. $H_{\text{TNN}}\left(\bm{k}\right)$ include all spin-independent terms up to the third-nearest-neighbor hopping
\begin{eqnarray}
H_{\text{TNN}}\left(\bm{k}\right)&=&
\begin{pmatrix}
V_0 & V_1 & V_2 \\ 
V_1^{\ast} & V_{11} & V_{12} \\ 
V_2^{\ast} & V_{12}^{\ast} & V_{22}
\end{pmatrix}.
\end{eqnarray}
The $\sigma_z \otimes L_z$-term accounts for the atomic spin-orbit coupling with $L_z$ being the orbital angular momentum operator
\begin{eqnarray}
 L_z&=&
 \begin{pmatrix}
0 & 0 & 0 \\ 
0 & 0 & -2i \\ 
0 & 2i & 0
\end{pmatrix}.
\end{eqnarray}
With the simplified notation $\left(\alpha, \beta\right) \equiv \left(\frac{1}{2}k_xa, \frac{\sqrt{3}}{2}k_ya\right)$, the expressions for the matrix elements in $H_{\text{TNN}}\left(\bm{k}\right)$ are listed below:
\begin{eqnarray}
V_0&=&\epsilon_1-\epsilon_0+2t_0\left(2\cos\alpha\cos\beta+\cos 2\alpha\right)+2r_0\left(2\cos3\alpha\cos\beta+\cos2\beta\right)+2u_0\left(2\cos2\alpha\cos2\beta+\cos4\alpha\right),\\
\text{Re}\left[V_1\right]&=&-2\sqrt{3}t_2\sin\alpha\sin\beta+2\left(r_1+r_2\right)\sin3\alpha\sin\beta-2\sqrt{3}u_2\sin2\alpha\sin2\beta, \\\nonumber
\text{Im}\left[V_1\right]&=&2t_1\sin\alpha\left(2\cos\alpha+\cos\beta\right)+2\left(r_1-r_2\right)\sin3\alpha\cos\beta+2u_1\sin2\alpha\left(2\cos2\alpha+\cos2\beta\right),\\
\text{Re}\left[V_2\right]&=&2t_2\left(\cos2\alpha-\cos\alpha\cos\beta\right)-\frac{2}{\sqrt{3}}\left(r_1+r_2\right)\left(\cos3\alpha\cos\beta-\cos2\beta\right)+2u_2\left(\cos4\alpha-\cos2\alpha\cos2\beta\right), \\\nonumber
\text{Im}\left[V_2\right]&=&2\sqrt{3}t_1\cos\alpha\sin\beta+\frac{2}{\sqrt{3}}\left(r_1-r_2\right)\sin\beta\left(\cos3\alpha+2\cos\beta\right)+2\sqrt{3}u_1\cos2\alpha\sin2\beta,\\
V_{11}&=&\epsilon_2-\epsilon_0+\left(t_{11}+3t_{22}\right)\cos\alpha\cos\beta+2t_{11}\cos2\alpha+4r_{11}\cos3\alpha\cos\beta+2\left(r_{11}+\sqrt{3}r_{12}\right)\cos2\beta\\\nonumber
&+&\left(u_{11}+3u_{22}\right)\cos2\alpha\cos2\beta+2u_{11}\cos4\alpha,\\
\text{Re}\left[V_{12}\right]&=&\sqrt{3}\left(t_{22}-t_{11}\right)\sin\alpha\sin\beta+4r_{12}\sin3\alpha\sin\beta+\sqrt{3}\left(u_{22}-u_{11}\right)\sin2\alpha\sin2\beta, \\\nonumber
\text{Im}\left[V_{12}\right]&=&4t_{12}\sin\alpha\left(\cos\alpha-\cos\beta\right)+4u_{12}\sin2\alpha\left(\cos2\alpha-\cos2\beta\right),\\
V_{22}&=&\epsilon_2-\epsilon_0+\left(3t_{11}+t_{22}\right)\cos\alpha\cos\beta+2t_{22}\cos2\alpha+2r_{11}\left(2\cos3\alpha\cos\beta+\cos2\beta\right)\\\nonumber
&+&\frac{2}{\sqrt{3}}r_{12}\left(4\cos3\alpha\cos\beta-\cos2\beta\right)+\left(3u_{11}+u_{22}\right)\cos2\alpha\cos2\beta+2u_{22}\cos4\alpha.
\end{eqnarray}
Parameters used in $H_{0}(\bm{k})$ are tabulated in Table \ref{table:MoS2}. Note that a constant term $\epsilon_{0} = 0.45$ eV is introduced in $V_0, V_{11}, V_{22}$ to ensure that the Fermi level lies above the maximum of the topmost valence band and all states with negative energies are occupied. 

The inter-layer coupling $T_{R}(\bm{k})$ is modelled by considering spin-preserving nearest-neighbor inter-layer tunneling processes: $T_{R}(\bm{k}) = \sigma_0 \otimes t_{R}(\bm{k})$. The $C_{3v}$ point group symmetry enforces $t_{R}(\bm{k})$ to take the form
\begin{eqnarray}\label{eq:interlayer3R}
t_{R}(\bm{k}) = 
\begin{pmatrix}
w_0 f_0(\bm{k}) & i \frac{\gamma_0}{\sqrt{2}} (f_+(\bm{k}) - f_-(\bm{k})) & \frac{\gamma_0}{\sqrt{2}} (f_+(\bm{k}) + f_-(\bm{k})) \\
i \frac{\gamma_1}{\sqrt{2}} (f_+(\bm{k}) - f_-(\bm{k})) & w_1 f_0(\bm{k}) & 0\\
\frac{\gamma_1}{\sqrt{2}} (f_+(\bm{k}) + f_-(\bm{k})) & 0 & w_1 f_0(\bm{k}) 
\end{pmatrix},
\end{eqnarray}
where the special functions in $t_{R}(\bm{k})$ are defined as
\begin{eqnarray}
f_0 (\bm{k}) &=& ( e^{i \bm{k} \cdot \bm{\delta}_1} +  e^{i \bm{k} \cdot \bm{\delta}_2} +   e^{i \bm{k} \cdot \bm{\delta}_3})/3, \\\nonumber
f_+ (\bm{k}) &=& (e^{i \bm{k} \cdot \bm{\delta}_1} + \omega_{+} e^{i \bm{k} \cdot \bm{\delta}_2} +  \omega_{-} e^{i \bm{k} \cdot \bm{\delta}_3})/3, \\\nonumber
f_- (\bm{k}) &=& (e^{i \bm{k} \cdot \bm{\delta}_1} + \omega_{-} e^{i \bm{k} \cdot \bm{\delta}_2} +  \omega_{+} e^{i \bm{k} \cdot \bm{\delta}_3})/3.
\end{eqnarray} 
Without loss of generality we consider here for AB stacking: $\bm{\delta}_1 = -\frac{a}{\sqrt{3}} \hat{y}$, and $\bm{\delta}_n = C^{(n-1)}_{3z} \bm{\delta}_1$. In terms of momentum $\bm{p} \equiv \bm{k} - \xi \bm{K}$ measured from $\pm K$ and up to first order in $\bm{p}$, the special functions in $t_{R}(\bm{k})$ are given by  
\begin{eqnarray}\label{eq:fproperty}
f_{0}(\bm{p} + \xi \bm{K}) &=& -\frac{a}{2\sqrt{3}}(\xi p_x - ip_y), \\\nonumber
f_{+}(+\bm{K} + \bm{p}) &=& i\frac{a}{2\sqrt{3}}(-p_x + ip_y), \\\nonumber
f_{+}(-\bm{K} + \bm{p}) &=& 1, \\\nonumber
f_{-}(+\bm{K} + \bm{p}) &=& 1, \\\nonumber
f_{-}(-\bm{K} + \bm{p}) &=& i\frac{a}{2\sqrt{3}}(p_x + ip_y).
\end{eqnarray}
The parameters used in the tight-binding model Eq.\ \eqref{eq:H3R} for bilayer 3R-MoS$_2$ are adapted from Ref.\ \cite{Yang2022S} and presented in Table \ref{table:MoS2}. We note that the  matrix elements in the 12-band tight-binding model in Eq.~\eqref{eq:H3R} includes not only direct $d-d$ electron hopping but also indirect hopping due to interactions between chalcogen $p$-orbitals and transition-metal $d$ orbitals (see paragraph following Eq.3 of Refs.~\cite{GuiBinS}).

\begin{table}[ht]
\caption{Tight-binding parameters for 3R bilayer MoS$_2$. All parameters are set in units of eV. Parameters are adapted previous DFT studies and experiments~\cite{GuiBinS,Yang2022S,HongyiS}.} 
\centering 
\begin{tabular}{c c c c c c c c c c} 
\hline\hline 
$\epsilon_1$ & $\epsilon_2$ & $t_0$ & $t_1$ & $t_2$ & $t_{11}$ & $t_{12}$ & $t_{22}$ & $r_0$ & $r_1$\\
$r_2$ & $r_{11}$ & $r_{12}$ & $u_0$ & $u_1$ & $u_2$ & $u_{11}$ & $u_{12}$ & $u_{22}$ & $\lambda$\\ 
$w_0$ & $w_1$ & $\gamma_0$ & $\gamma_1$ \\[0.5ex] 
\hline 
0.813 & 1.707 & -0.146 & -0.114 & 0.506 & 0.085 & 0.162 & 0.073 & 0.060 & -0.236 \\ 
0.067 & 0.016 & 0.087 & -0.038 & 0.046 & 0.001 & 0.266 & -0.176 & -0.150 & 0.073 \\ 
-0.4 & -0.02 & -0.118 & 0.118 \\
[1ex] 
\hline 
\end{tabular}
\label{table:MoS2} 
\end{table}

\subsubsection*{B. Effective SSH model at $\pm K$}

As explained in Refs.\ \cite{Yang2022S,HongyiS}, conduction band state and valence band state at the three-fold ($\mathcal{C}_3$)-invariant $K$-points in monolayer MoS$_2$ have different $\mathcal{C}_3$-angular momentum quantum numbers $m_z = 0$ and $m_z = -1$ respectively. In the 3R-stacking configuration, however, the relative lateral shift between two layers modifies the angular momentum quantum numbers at the $K$-points. For the AB-stacking configuration considered here, the states at $+K$ in layer 2 acquire an extra angular momentum of +1 from the vantage point of layer 1. As a result, the conduction band state $\ket{c, +K, 1}$ in layer 1 and valence band state $\ket{v, +K, 2}$ in layer 2 share the same $m_z = 0$, while the conduction band state $\ket{c, +K, 2}$ in layer 2 has $m_z = +1$ and the valence band state $\ket{v, +K, 1}$ in layer 1 has $m_z = -1$ (see Table \ref{table:00}). Therefore, the inter-layer coupling at $+K$ is non-vanishing only between $\ket{c, +K, 1}$ and $\ket{v, +K, 2}$, while $\ket{c, +K, 2}$ and $\ket{v, +K, 1}$ remain decoupled, as we show schematically in Fig.\ 2b of the main text. 

We note that the asymmetric inter-layer coupling above is well-captured by the symmetry-based tight-binding model for 3R-bilayer TMD we present in subsection A above. Note that for TMDs the conduction band edge at $\pm K$ are dominated by the $d_{z^2}$-orbitals while the valence band edge at $\pm K$ originate from $\ket{d_{\pm}} \equiv \ket{d_{x^2-y^2} \pm i d_{xy}}$, respectively~\cite{GuiBinS,DiXiaoS}. This motivates us to rewrite $t_{R}(\bm{k})$ in the new basis $\{ \ket{d_{z^2}}, \ket{d_{+}}, \ket{d_{-}} \}$, which amounts to the basis transformation: $\tilde{t}_{R}(\bm{k}) = U t_{R}(\bm{k}) U^{\dagger}$, where
\begin{eqnarray}
\tilde{t}_R(\bm{k}) = 
\begin{pmatrix}
w_0 f_0(\bm{k}) & \gamma_0 f_{-}(\bm{k}) & \gamma_0 f_{+}(\bm{k})\\
\gamma_1 f_{+}(\bm{k}) & w_1 f_{0}(\bm{k}) & 0\\
\gamma_1 f_{-}(\bm{k}) & 0 & w_1 f_{0}(\bm{k})    
\end{pmatrix}, \ \ \ 
& U = 
\begin{pmatrix}
1 & 0 & 0\\
0 & \frac{-i}{\sqrt{2}} & \frac{1}{\sqrt{2}}\\
0 & \frac{i}{\sqrt{2}} & \frac{1}{\sqrt{2}}\\
\end{pmatrix}.
\end{eqnarray}

Given the properties of the special functions in Eq.\ \eqref{eq:fproperty}, it is straightforward to show that in the basis $\{ \ket{v, \pm K, 1}, \ket{c, \pm K, 1}, \ket{v, \pm K, 2}, \ket{c, \pm K, 2} \}$, the effective Hamiltonian at $\pm K$ can be written as
\begin{eqnarray}\label{eq:Heff_mos2}
H_{{\rm eff}, \xi}(\bm{p}) = 
\begin{pmatrix}
E_{v1} & v_0(\xi p_x + i p_y) & u_1(\xi p_x - i p_y) & i v_1 (-\xi p_x -ip_y)\\
v_0(\xi p_x - i p_y) & E_{c1} & \gamma_0 & u_0(\xi p_x - i p_y)\\
u_1(\xi p_x + i p_y) & \gamma_0 & E_{v2} & v_0(\xi p_x + i p_y)\\
-i v_1 (-\xi p_x + ip_y) & u_0(\xi p_x + i p_y) & v_0(\xi p_x - i p_y) & E_{c2}
\end{pmatrix}.
\end{eqnarray}
Here, $\xi = \pm$ is the valley index, $E_{c1} = E_{c2} = E_c$ and $E_{v1} = E_{v2} = E_v$ denote the energies at conduction and valence band edges in each decoupled monolayer. $v_0$ denotes the Dirac velocity from intra-layer hopping~\cite{DiXiaoS}, $u_1 = -\frac{w_1 a}{2 \sqrt{3}}$, $u_0 = -\frac{w_0 a}{2 \sqrt{3}}$ are the velocity terms from inter-layer intra-orbital hopping, and $\gamma_0$ denotes the direct coupling between $c_1$ and $v_2$ states with the same $m_z = 0$. Clearly, Eq.~\eqref{eq:Heff_mos2} is formally the same as Eq.\ (4) of the main text for AB-stacked honeycomb bilayers. For momentum near $\pm K$ with $\bm{p} \simeq 0$, the inter-layer coupling only occurs between $\ket{c, \pm K, 1}$ and $\ket{v, \pm K, 2}$ and Eq.\ \eqref{eq:Heff_mos2} reduces to the same form as Eq.\ref{eq:heff_berry}, thus supporting a polar Berry phase of the form in Eq.\ (3) of the main text.

\begin{table}[ht]
\setlength{\tabcolsep}{10pt}
\caption{Values of $\mathcal{C}_3$-angular momentum quantum number $m_z$ for states at $+K$ in AB-stacked 3R bilayer TMD. } 
\centering 
\begin{tabular}{c c} 
\hline\hline
State at +K  & $m_z$ \\
\hline\hline
$\ket{c,+K,1}$ &  0   \\
$\ket{v,+K,1}$  &  $-1$ \\
$\ket{c,+K,2}$ &  $+1$ \\
$\ket{v,+K,2}$ &  0  \\
\hline\hline
\end{tabular}
\label{table:00} 
\end{table}

\subsection*{3. Bilayer T$_{\rm d}$-MX$_2$}

Here we present microscopic models describing bilayer transition-metal dichalcogenides (TMDs) in the T$_{\rm d}$-phase with general chemical formula MX$_2$ (M = Mo, W; X = Te). We first introduce a symmetry-based $\bm{k}\cdot\bm{p}$ model near the $\Gamma$-point for bilayer T$_{\rm d}$ MX$_2$, and then demonstrate how the $\bm{k}\cdot\bm{p}$ model leads to effective massive Dirac Hamiltonians near Q-points, which can be mapped to the spinful SSH model described in the main text. Finally, we regularize the $\bm{k}\cdot\bm{p}$ model to obtain a microscopic lattice model on a rectangular lattice, which allows us to perform the self-consistent calculation of $P_z$ using the expression given in Eq.\ (6) of the main text. 

\subsubsection*{A. \(\mathbf{k \cdot p}\) Hamiltonian}

As we mentioned in the main text, the bilayer T$_{\rm d}$-structure is formed by stacking two centrosymmetric 1T'-monolayers which are related to each other through a $\mathcal{C}_{2z}$ rotation followed by a lateral shift of half the unit cell along the $y$-direction: $(x,y) \mapsto (x,y+b/2)$~\cite{CarS}. For each isolated 1T'-monolayer, the relevant low-energy physics and nontrivial band topology can be captured by a four-band $\bm{k}\cdot\bm{p}$ Hamiltonian in the basis of symmetrized $d, p$-orbitals ($\ket{p_x, \uparrow}, \ket{p_x, \downarrow}, \ket{d_{xz}, \uparrow}, \ket{d_{xz}, \downarrow}$) \cite{JunweiS, BenjaminS}: 
\begin{eqnarray}\label{eq:H0}
H_{0}(\bm{k}) &=& 
\begin{pmatrix}
E_p(\bm{k})           & 0                      & -i v k_y + A_z k_x & A_x k_y - i A_y k_x \\
0                     & E_p(\bm{k})            & A_x k_y + i A_y k_x & -i v k_y - A_z k_x \\
i v k_y + A_z k_x & A_x k_y - i A_y k_x        & E_d(\bm{k})           & 0 \\
A_x k_y + i A_y k_x & i v k_y - A_z k_x        & 0                     & E_d(\bm{k}) 
\end{pmatrix}\\\nonumber
&=& E_{+}(\bm{k}) s_0 \sigma_0 + E_{-}(\bm{k}) s_z \sigma_0 + v k_y s_y \sigma_0 + A_x k_y s_x \sigma_x + A_y k_x  s_x \sigma_y + A_z k_x s_x \sigma_z,
\end{eqnarray}
where $E_{p}(\bm{k}) = -t_{xp} k_x^2 - t_{yp} k_y^2 - \mu_p$, $E_{d}(\bm{k}) = -t_{xd} k_x^2 - t_{yd} k_y^2 + t'_x k_x^4 + t'_y k_y^4 - \mu_d$, $E_{+}(\bm{k}) = [E_p(\bm{k}) + E_d(\bm{k})]/2$, $E_{-}(\bm{k}) = [E_p(\bm{k}) - E_d(\bm{k})]/2$. $A_{\alpha = x,y,z}$ characterizes the strength of spin-orbit coupling (SOC) which pins the spins to $\alpha$-direction and induces nontrivial topological gaps at  crossing points between SOC-free energy bands $E_p(\bm{k})$ from $p$-orbitals and $E_d(\bm{k})$ from $d$-orbitals \cite{BenjaminS}. $s_{\alpha = x,y,z}$ ($\sigma_{\alpha = x,y,z}$) are Pauli matrices operating on the orbital (spin) subspace. The symmetry generators of 1T'-monolayer consist of spatial inversion $\mathcal{I}$ and an in-plane mirror reflection $\mathcal{M}_x$. They transform the $p_x,d_{xz}$-orbitals as $\mathcal{I} \ket{p_x, \sigma} = - \ket{p_x, \sigma}$, $\mathcal{I} \ket{d_{xz}, \sigma} = + \ket{d_{xz}, \sigma}$, $\mathcal{M}_x \ket{p_x, \uparrow} = -i \ket{p_x, \downarrow}$, $\mathcal{M}_x \ket{d_{xz}, \uparrow} = -i \ket{d_{xz}, \downarrow}$ and thus the matrix representations of $\mathcal{I}, \mathcal{M}_x$ can be written as $U(\mathcal{I}) = s_z \sigma_0$, $U(\mathcal{M}_x) = -i s_0 \sigma_x$. The Hamiltonian $H_0$ in Eq.~\eqref{eq:H0} satisfies the symmetry conditions: $U(\mathcal{I})H_0(\bm{k}) U^{-1}(\mathcal{I}) = H_0(-\bm{k})$, $U(\mathcal{M}_x)H_0(k_x, k_y) U^{-1}(\mathcal{M}_x) = H_0(-k_x, k_y)$.

Without loss of generality, we choose the top 1T'-monolayer as layer 1 and set $H_{1} \equiv H_0$. The Hamiltonian for the bottom layer (labeled as layer 2) can then be derived by applying to $H_1$ a $\mathcal{C}_{2z}$-rotation. In the basis of $H_0$ the rotation is represented by $U(\mathcal{C}_{2z}) = -is_0\sigma_z$,= and we thus find
\begin{eqnarray}
H_2(\bm{k}) &=&   U(\mathcal{C}_{2z}) H_1(-\bm{k}) U^{-1}(\mathcal{C}_{2z})\\\nonumber
&=& E_{+}(\bm{k}) s_0 \sigma_0 + E_{-}(\bm{k}) s_z \sigma_0 - v k_y s_y \sigma_0 + A_x k_y s_x \sigma_x + A_y k_x  s_x \sigma_y - A_z k_x s_x \sigma_z.
\end{eqnarray}

For the inter-layer coupling Hamiltonian $H_{T}$ between basis states in $H_1$ and $H_2$, we consider the leading-order spin-preserving intra-cell terms: $g_{mn, \sigma\sigma'} = \braket{1, m|H_{T}|2, n} \delta_{\sigma\sigma'}$, with $m,n = \{p_x,d_{xz}\}$ and $\sigma, \sigma' = \uparrow, \downarrow$. Including $H_{T}$, the total $\bm{k}\cdot\bm{p}$ Hamiltonian for bilayer T$_{\rm d}$-MX$_2$ can be written as
\begin{eqnarray}
H(\bm{k}) = 
\begin{pmatrix}
H_1(\bm{k}) & H_{T}\\
H_T^{\dagger} & H_2(\bm{k})
\end{pmatrix}, &&
H_{T} = 
\begin{pmatrix}
 g_{pp} & g_{pd}\\
 g_{dp} & g_{dd}
\end{pmatrix}
\otimes \sigma_0.
\end{eqnarray}
We note that due to the low symmetry of the bilayer T$_{\rm d}$ phase, there are essentially no constraints from point group symmetries on the entries $g_{mn}$ in $H_T$, while time-reversal $\mathcal{T}$ requires $H_T = i\sigma_y\mathcal{K} H_T (-i\sigma_y\mathcal{K}) = H^{\ast}_T$, which implies all $g_{mn}$ are real. By further introducing Pauli matrices $\tau_{\alpha=x,y,z}$ acting on the layer subspace, we can rewrite the total bilayer $\bm{k}\cdot\bm{p}$ Hamiltonian as
\begin{eqnarray}\label{eq:Hkdotp}
H(\bm{k}) &=& E_{+}(\bm{k}) \tau_0 s_0 \sigma_0 + E_{-}(\bm{k}) \tau_0 s_z \sigma_0 + v k_y \tau_z s_y \sigma_0 + A_x k_y \tau_0 s_x \sigma_x + A_y k_x \tau_0 s_x \sigma_y + A_z k_x \tau_z s_x \sigma_z \\\nonumber
&+& g_{0,+} \tau_x s_0 \sigma_0 + g_{0,-} \tau_x s_z \sigma_0 + g_{1,+} \tau_x s_x \sigma_0 +  g_{1,-} \tau_y s_y \sigma_0,
\end{eqnarray}
where $g_{0,+} = (g_{pp}+g_{dd})/2$, $g_{0,-} = (g_{pp}-g_{dd})/2$, $g_{1,+} = (g_{dp}+g_{pd})/2$, $g_{1,-} = (g_{dp}-g_{pd})/2$. 

In the main text, we mentioned that inversion is broken by the $g_{1,+}$-term, while $g_{0,-}, g_{1,-}$-terms preserve inversion symmetry. Here, we derive the parities of the inter-layer $g_{0,+}, g_{1,+}, g_{0,-}, g_{1,-}$-terms under spatial inversion $\mathcal{I}$. We note that for a bilayer system in general, the inversion operation $\mathcal{I}$ can be decomposed as a swap between the two layers followed by a 2D inversion operation in each layer. Therefore, $\mathcal{I}$ transforms the basis states of $H(\bm{k})$ in Eq.~\eqref{eq:Hkdotp} as: $\mathcal{I} \ket{1, p_x, \sigma} = - \ket{2, p_x, \sigma}$, $\mathcal{I} \ket{1, d_{xz}, \sigma} = + \ket{2, d_{xz}, \sigma}$, where $\sigma = \uparrow, \downarrow$, and the matrix representation of $\mathcal{I}$ is given by $U(\mathcal{I}) = -\tau_x s_z \sigma_0$. Making use of the anti-commutation relations of Pauli matrices, it is straightforward to show that
\begin{eqnarray}\label{eq:InterlayerSymmetry}
U(\mathcal{I}) \tau_x s_0 \sigma_0  U^{-1}(\mathcal{I}) &=&  \tau_x s_0 \sigma_0,\\\nonumber
U(\mathcal{I}) \tau_x s_z \sigma_0  U^{-1}(\mathcal{I}) &=&  \tau_x s_z \sigma_0,\\\nonumber
U(\mathcal{I}) \tau_x s_x \sigma_0  U^{-1}(\mathcal{I}) &=&  -\tau_x s_x \sigma_0,\\\nonumber
U(\mathcal{I}) \tau_y s_y \sigma_0  U^{-1}(\mathcal{I}) &=&  \tau_y s_y \sigma_0,\\\nonumber
\end{eqnarray}
thus the $g_{1,+}$-term is the only inversion-breaking term, while all other terms are compatible with inversion symmetry.

\subsubsection*{B. Tight-binding model}

Here, we present a tight-binding model for bilayer T$_{\rm d}$-MX$_2$ which extends the effective model near $Q, Q'$ introduced above to the entire Brillouin zone. The tight-binding model for monolayer WTe$_2$ is given by
\begin{equation}\label{eq:monolayerWTe2}
    H_{\text{TB}}(\bm{k})=
    \begin{pmatrix}
        E_p(\bm{k})-\mu & 0 & -iv_0\sin(k_y b)+\alpha_z\sin(k_x a) & -i \alpha_y\sin(k_x a)+\alpha_x\sin(k_y b)\\
        & E_p(\bm{k})-\mu & i\alpha_y \sin(k_x a)+\alpha_x\sin(k_y b) & -i v_0\sin(k_y b) - \alpha_z \sin(k_x a)\\
        & & E_d(\bm{k})-\mu & 0 \\
        h. c. & & & E_d(\bm{k})-\mu
    \end{pmatrix}
\end{equation}
where the lower triangle is related to the upper one by Hermitian conjugate, and
\begin{align}
E_p(\bm{k})&=2 t_{1p}\cos(k_x a)+2t_{2p}\cos(k_y b)-u_p-2(t_{1p}+t_{2p})\\\nonumber
E_d(\bm{k})&=2 t_{1d}\cos(k_x a)+2t_{2d}\cos(k_y b)+2t_{1d}' \cos(2k_xa)-u_d-2(t_{1d}+t_{2d}+t_{1d}').\nonumber     
\end{align}

Bilayer T$_{\rm d}$-WTe$_2$ structure is formed by stacking two monolayers related  each other by $C_{2z}=M_z\mathcal{I}$ with a lateral shift $(x,y)\rightarrow(x,y+\frac{b}{2})$~\cite{CarS}.  The two layers are related by a mirror symmetry as discussed above. If the tight-binding Hamiltonian in layer 1 is given by $H(\bm{k},v,\alpha_{z})$, the symmetry considerations imply that the Hamiltonian in layer 2 will be $H(\bm{k},-v,-\alpha_{z})$. In the basis, $[\vert 1, p , \uparrow\rangle,\vert 1, p , \downarrow\rangle, \vert 1, d , \uparrow\rangle, \vert 1, d , \downarrow\rangle, \vert 2, p , \uparrow\rangle,\vert 2, p , \downarrow\rangle, \vert 2, d , \uparrow\rangle, \vert 2, d , \downarrow\rangle]$, the tight-binding Hamiltonian for T$_{\rm d}$ bilayer is given by
\begin{equation}\label{eq:tb_tmd}
    H_{T_{\rm d}}(\bm{k}) =
    \begin{pmatrix}
        H^{(1)}(\bm{k},v,\alpha_z) & H_T\\
        H_T^{\dagger} & H^{(2)}(\bm{k},-v,-\alpha_z)
    \end{pmatrix},
\end{equation}
where the inter-layer coupling matrix is given by
\begin{equation}
    H_T=
    \begin{pmatrix}
        g_{pp} & 0 & g_{pd} & 0\\
        0 & g_{pp} & 0 & g_{pd}\\
        g_{dp} & 0 & g_{dd} & 0\\
        0 & g_{dp} & 0 & g_{dd}
    \end{pmatrix}.
\end{equation}

The $\mathbf{k}\cdot\mathbf{p}$ model parameters and the corresponding tight-binding model parameters for each monolayer and the interlayer coupling constants are provided in Tables \ref{tab:td_s1} and \ref{tab:td_s2} respectively. 

\begin{table}[]
\caption{\label{tab:td_s1} Parameters of the $\mathbf{k}\cdot\mathbf{p}$ and the tight-binding model for monolayer T$_{\rm d}$-WTe$_{2}$ adapted from Ref.~\cite{BenjaminS}. Lattice constants: $a=3.49$ {\AA}, $b=6.31$ {\AA}.}
\begin{tabular}{ll||lll}
\hline\hline
\multicolumn{2}{l||}{$\mathbf{k}\cdot\mathbf{p}$  \textbf{parameters}}                                & \multicolumn{3}{l}{\textbf{Tight-binding parameters}}                                                                                                                                \\ \hline
\multicolumn{1}{l}{Parameter}                              & Value  & \multicolumn{1}{l}{Parameter}  & \multicolumn{1}{l}{Relation to $\mathbf{k}\cdot\mathbf{p}$ model}                                                & Value                                              \\ \hline
\multicolumn{1}{l}{$t_{xp}$ (eV$\cdot$\AA$^2$)}   & 18.477 & \multicolumn{1}{l}{$(a,b)$}    & \multicolumn{1}{l}{-}                                                                  & (3.49 \AA, 6.31 \AA) \\ 
\multicolumn{1}{l}{$t_{yp}$ (eV$\cdot$\AA$^2$)} & 24.925 & \multicolumn{1}{l}{$t_{1p}$}   & \multicolumn{1}{l}{$\frac{t_{xp}}{a^2}$}                                               & 1.517                                              \\ 
\multicolumn{1}{l}{$\mu_p$ (eV)}                           & -1.39  & \multicolumn{1}{l}{$t_{2p}$}   & \multicolumn{1}{l}{$\frac{t_{yp}}{b^2}$}                                               & 0.626                                              \\ 
\multicolumn{1}{l}{$t_{xd}$ (eV$\cdot$\AA$^2$)}   & 2.594  & \multicolumn{1}{l}{$u_p$}      & \multicolumn{1}{l}{$\mu_p$}                                                            & -1.39                                              \\ 
\multicolumn{1}{l}{$t_{yd}$ (eV$\cdot$\AA$^2$)}   & -2.46  & \multicolumn{1}{l}{$t_{1d}$}   & \multicolumn{1}{l}{$\frac{1}{2}\left(\frac{t_{xd}}{a^2}+\frac{t_{x}'.12}{a^4}\right)$} & -0.387                                             \\ 
\multicolumn{1}{l}{$t_{x}'$ (eV$\cdot$\AA$^4$)}   & 24.887 & \multicolumn{1}{l}{$t_{2d}$}   & \multicolumn{1}{l}{$\frac{1}{3}\left(\frac{4t_{yd}}{b^2}-\frac{12 t_y'}{b^4}\right)$}  & -0.0619                                            \\ 
\multicolumn{1}{l}{$t_{y}'$ (eV$\cdot$\AA$^4$)}   & -8.191 & \multicolumn{1}{l}{$t_{2d}'$}  & \multicolumn{1}{l}{$\frac{1}{12}\left(\frac{12t_x'}{b^4}-\frac{t_{yd}}{b^2}\right)$}   & 0.15                                               \\ 
\multicolumn{1}{l}{$\mu_d$ (eV)}                           & 0.062  & \multicolumn{1}{l}{$u_d$}      & \multicolumn{1}{l}{$\mu_d$}                                                            & 0.062                                              \\ 
\multicolumn{1}{l}{$v$ (eV$\cdot$\AA)}            & 2.34   & \multicolumn{1}{l}{$v$}        & \multicolumn{1}{l}{$v/b$}                                                              & 0.371                                              \\ 
\multicolumn{1}{l}{$A_x$ (eV$\cdot$\AA)}          & 0.17   & \multicolumn{1}{l}{$\alpha_x$} & \multicolumn{1}{l}{$\alpha_x/b$}                                                       & 0.027                                              \\ 
\multicolumn{1}{l}{$A_y$ (eV$\cdot$\AA)}          & 0.57   & \multicolumn{1}{l}{$\alpha_y$} & \multicolumn{1}{l}{$\alpha_y/a$}                                                       & 0.163                                              \\ 
\multicolumn{1}{l}{$A_z$ (eV$\cdot$\AA)}          & 0.07   & \multicolumn{1}{l}{$\alpha_z$} & \multicolumn{1}{l}{$\alpha_z/a$}                                                       & 0.02                                               \\ \hline\hline
\end{tabular}

\end{table}

\begin{table}[]
\caption{\label{tab:td_s2} Interlayer coupling parameters in units of eV for bilayer T$_{\rm d}$-WTe$_{2}$. The parameters are chosen to reproduce the band structures near $Q,Q'$ featuring coupled massive Dirac fermions~\cite{QiongS, DuS}.}
\centering
\begin{tabular}{llll}\hline\hline
$g_{pp}$  & $g_{dd}$  & $g_{pd}$   & $g_{dp}$ \\\hline
0.02 & 0.03 & 0.065 & 0  \\\hline\hline
\end{tabular}
\end{table}

\subsubsection*{C. Effective spinful SSH model and Berry phase at $Q, Q'$}

The bulk energy gaps in 1T'-monolayer and T$_{\rm d}$-bilayer are known to arise from mass terms that gap out the crossings between $d, p$-bands at the two inequivalent $Q$ and $Q'$ points, with $\bm{Q} = (Q_x, 0)$, $\bm{Q}' = (-Q_x, 0)$ \cite{JunweiS, BenjaminS, QiongS, DuS}. In the following, we present a detailed derivation of the effective massive Dirac Hamiltonians from the $\bm{k}\cdot\bm{p}$ Hamiltonian in Eq.~\eqref{eq:Hkdotp}. We focus on $\bm{Q} = (Q_x, 0)$ first, and the physics at $-\bm{Q}$ follows from time-reversal symmetry.

First, we keep only the leading-order SOC terms at $Q$: $H_{\rm SOC}(\bm{k} \simeq \bm{Q}) \simeq m_x \tau_0 s_x \sigma_y + m_y \tau_0 s_x \sigma_y + m_z \tau_z s_x \sigma_z$, where $m_{\alpha = x,y,z} \equiv A_{\alpha=x,y,z} Q_x$. Next, near the band crossing points we have $E_{-}(\bm{k} \simeq \bm{Q}) = u p_x + o(p_x^2, p_y^2)$, where the momentum $\bm{p} = (p_x, p_y) = (k_x - Q_x, k_y) = \bm{k} - \bm{Q}$ is now measured from $\bm{Q} = (Q_x, 0)$. These simplifications for Eq.~\eqref{eq:Hkdotp} allow us to write down an effective Hamiltonian for the $Q$-point:
\begin{eqnarray}\label{eq:Hsimplified}
H_{Q}(\bm{p}) &=& E_{+}(\bm{Q}) \tau_0 s_0 \sigma_0 + u p_x \tau_0 s_z \sigma_0 + v p_y \tau_z s_y \sigma_0 + m_x \tau_0 s_x \sigma_y + m_y \tau_0 s_x \sigma_y + m_z \tau_z s_x \sigma_z\\\nonumber
&+& g_{0,+} \tau_x s_0 \sigma_0 + g_{0,-} \tau_x s_z \sigma_0 + g_{1,+} \tau_x s_x \sigma_0 +  g_{1,-} \tau_y s_y \sigma_0.
\end{eqnarray}

To derive massive Dirac Hamiltonians based on Eq.~\eqref{eq:Hsimplified}, we note that the energy scale of the intra-layer SOC term is of order 150 meV \cite{JunweiS}, which strongly dominates over other terms for $\bm{k} \simeq \bm{Q}$. This motivates us to perform a basis transformation $V_{l}$ on each layer $l=1,2$, which diagonalizes the SOC terms $H^{l=1,2}_{\rm SOC}$ on each layer, such that $H_{Q}(\bm{p})$ in Eq.~\eqref{eq:Hsimplified} is recast as $\tilde{H}_Q(\bm{p}) = V^{\dagger} H_{Q}(\bm{p}) V $ with the transformation matrices given by
\begin{eqnarray}\label{eq:basistrans}
V &=& 
\begin{pmatrix}
V_{1} & 0_{4 \times 4}\\
0_{4 \times 4} & V_{2}
\end{pmatrix},\\
V_{l} &=& \frac{1}{\sqrt{2}} 
\begin{pmatrix}
1 & 1\\
1 & -1
\end{pmatrix}
\otimes
\begin{pmatrix}
e^{i\phi/2}\cos(\theta_l/2) & e^{i\phi/2}\sin(\theta_l/2)\\
e^{-i\phi/2}\sin(\theta_l/2) & -e^{-i\phi/2}\cos(\theta_l/2)
\end{pmatrix}.
\end{eqnarray}
Here $\phi = \tan^{-1}(m_{y}/m_{x})$ is the azimuthal angle of the SOC vector, $\theta_l = \cos^{-1}[(-1)^{l+1}m_{z}/m]$ is the polar angle of the SOC vector, and $m \equiv \sqrt{m^2_x + m^2_y + m^2_z}$ is the total magnitude of the SOC term. Note that the layer-dependent sign in $\theta_l$ arises from the different signs in the $m_z$-term between layer 1 and 2 (indicated by $\tau_z$ in Eq.~\eqref{eq:Hsimplified}); thus $\theta_{1}, \theta_{2}$ are related by $\theta_{2} = \pi - \theta_1$ which implies $\cos(\theta_2/2) = \sin(\theta_1/2), \sin(\theta_2/2) = \cos(\theta_1/2)$. The transformed effective Hamiltonian in the basis of the eigenstates of $H_{\rm SOC}$ takes the form:
\begin{eqnarray}\label{HQp}
\tilde{H}_Q(\bm{p}) &=& E_{+}(\bm{Q}) I_{8 \times 8} +
\begin{pmatrix}
\tilde{H}_{1, Q}(\bm{p}) & \hat{g}(\theta_1)\\
\hat{g}^{\dagger}(\theta_1) & \tilde{H}_{2, Q}(\bm{p})
\end{pmatrix},\\
\tilde{H}_{1, Q}(\bm{p}) &=& 
\begin{pmatrix}
m & 0 & up_x +ivp_y & 0\\
0 & -m & 0 & up_x +ivp_y \\
up_x -ivp_y & 0 & -m & 0\\
0 & up_x -ivp_y & 0 & m
\end{pmatrix},\\
\tilde{H}_{2, Q}(\bm{p}) &=& 
\begin{pmatrix}
m & 0 & up_x -ivp_y & 0\\
0 & -m & 0 & up_x -ivp_y \\
up_x +ivp_y & 0 & -m & 0\\
0 & up_x +ivp_y & 0 & m
\end{pmatrix},\\
\hat{g}(\theta_1) &=&
\begin{pmatrix}
g_{0,+} + g_{1,+} & g_{0,-} + g_{1,-}\\
g_{0,-} - g_{1,-} & g_{0,+} - g_{1,+}\\ 
\end{pmatrix}
\otimes 
\begin{pmatrix}
\sin(\theta_1) & \cos(\theta_1)\\
-\cos(\theta_1) & \sin(\theta_1)
\end{pmatrix},
\end{eqnarray}
where in reaching the form of $\hat{g}(\theta_1)$ we made use of
\begin{eqnarray}
\cos(\theta_2/2)\cos(\theta_1/2) + \sin(\theta_2/2)\sin(\theta_1/2) &=& \sin(\theta_1/2)\cos(\theta_1/2) + \cos(\theta_1/2)\sin(\theta_1/2) = \sin(\theta_1),\\\nonumber
\sin(\theta_2/2)\cos(\theta_1/2) - \cos(\theta_2/2)\sin(\theta_1/2) &=& \cos(\theta_1/2)\cos(\theta_1/2) - \sin(\theta_1/2)\sin(\theta_1/2) = \cos(\theta_1).
\end{eqnarray}

To further simplify $\tilde{H}_Q$, we note that the SOC in WTe$_2$ was shown to exhibit strong anisotropy with $\sqrt{m^2_x + m^2_y}\gg m_z$ \cite{BenjaminS}, which implies $|\cos(\theta_1)|=m_z/m\ll 1$, $\sin(\theta_1) \simeq 1$ and the inter-layer terms involving the $\cos(\theta_1)$ factor can be treated as weak perturbations. Setting $\lambda \equiv \cos(\theta_1)$, we can obtain a simplified effective Hamiltonian using straightforward second-order perturbation theory. The resulting effective Hamiltonian $\tilde{H}_{Q,\rm{eff}}$ has the same form as Eq.\ \eqref{HQp} with the blocks replaced by
\begin{eqnarray}\label{eq:effectiveHatQ}
\tilde{H}_{1, Q, \rm{eff}}(\bm{p}) &=& 
\begin{pmatrix}
\tilde{m}+V & 0 & up_x +ivp_y & 0\\
0 & -\tilde{m}-V & 0 & up_x +ivp_y \\
up_x -ivp_y & 0 & -\tilde{m} + V  & 0\\
0 & up_x -ivp_y & 0 & \tilde{m} - V
\end{pmatrix},\\
\tilde{H}_{2, Q, \rm{eff}}(\bm{p}) &=& 
\begin{pmatrix}
\tilde{m}+V & 0 & up_x -ivp_y & 0\\
0 & -\tilde{m}-V & 0 & up_x -ivp_y \\
up_x +ivp_y & 0 & -\tilde{m}+V & 0\\
0 & up_x +ivp_y & 0 & \tilde{m}-V
\end{pmatrix},\\
\hat{g}_{\rm eff} &=&
\begin{pmatrix}
\tilde{g}_{0,+} + \tilde{g}_{1,+} & g_{0,-} + g_{1,-}\\
g_{0,-} - g_{1,-} & \tilde{g}_{0,+} - \tilde{g}_{1,+}\\ 
\end{pmatrix}
\otimes 
\begin{pmatrix}
1 & 0\\
0 & 1
\end{pmatrix},
\end{eqnarray}  
where $\tilde{m} = m + \frac{\lambda^2(g_{0,+}^2+g_{1,+}^2)}{m}$ , $V = \frac{2\lambda^2 g_{0,+}g_{1,+}}{m}$, $\tilde{g}_{0,+} = g_{0,+} + \frac{\lambda^2g^2_{0,-}}{2g_{0,+}}$, $\tilde{g}_{1,+} = g_{1,+} + \frac{\lambda^2g^2_{1,-}}{2g_{1,+}}$. In the derivation of Eq.~\eqref{eq:effectiveHatQ} we have kept leading order perturbation terms only and dropped all the higher order  terms of the form $\lambda^2|g_{0,+}\pm g_{1,+}|^3/m^2$ that are negligibly small. It is worth noting that $V = \frac{2\lambda^2 g_{0,+}g_{1,+}}{m}$ originates from the combined effect of: (i) the $m_z$-SOC term in Eq.~\eqref{eq:Hsimplified}, which gives rise to a nonzero $\lambda \equiv \cos(\theta_1)$, and (ii) the $\mathcal{I}$-breaking inter-layer $g_{1,+}$-term in Eq.~\eqref{eq:InterlayerSymmetry}. It is evident from Eq.~\eqref{eq:effectiveHatQ} that the presence of the $V$-term lifts the two-fold spin degeneracy in each band, which manifests the broken $\mathcal{I}$-symmetry due to the $g_{1,+}$-term in Eq.~\eqref{eq:InterlayerSymmetry}.

We note that the form of \eqref{eq:effectiveHatQ} is now block-diagonalized into two effective independent spin sectors, which we define as $\tilde{\sigma} = \pm$, and $\tilde{H}_{Q,\rm{eff}}$ can be written in a more compact form as
\begin{eqnarray}
\tilde{H}_{Q,\rm{eff,\tilde{\sigma}}}(\bm{p}) &\simeq& E_{+}(\bm{Q}) I_{4 \times 4} + 
\begin{pmatrix}
\tilde{\sigma}\tilde{m}+\tilde{\sigma}V & up_x +ivp_y & \tilde{g}_{0,+} + \tilde{g}_{1,+} & g_{0,-} + g_{1,-}\\
up_x -ivp_y & -\tilde{\sigma}\tilde{m}+\tilde{\sigma}V & g_{0,-} - g_{1,-} & \tilde{g}_{0,+} - \tilde{g}_{1,+}\\
\tilde{g}_{0,+} + \tilde{g}_{1,+} & g_{0,-} - g_{1,-} & \tilde{\sigma}\tilde{m} + \tilde{\sigma}V  & up_x -ivp_y\\
g_{0,-} + g_{1,-} & \tilde{g}_{0,+} - \tilde{g}_{1,+} & up_x +ivp_y &  -\tilde{\sigma}\tilde{m} +  \tilde{\sigma}V
\end{pmatrix}.
\end{eqnarray}

 The physics at $-\bm{Q} = (-Q_x,0)$ follows from time-reversal symmetry $\mathcal{T}$: $\tilde{H}_{-Q}(\bm{p}) = \mathcal{T}\tilde{H}_{Q}(-\bm{p})\mathcal{T}^{-1}$, which amounts to the following changes $\tilde{\sigma} \mapsto -\tilde{\sigma}$, $\bm{p} \mapsto -\bm{p}$, $i \mapsto -i$ in $\tilde{H}_{Q}$.  Thus, the total effective Hamiltonian including $\xi =\pm$ for both $+\bm{Q}, -\bm{Q}$ valleys is written as
 \begin{eqnarray}\label{eq:CoupledMassiveDirac}
 \tilde{H}_{\xi Q,\rm{eff,\tilde{\sigma}}}(\bm{p}) &\simeq& E_{+}(\xi\bm{Q}) I_{4 \times 4} + 
\begin{pmatrix}
\xi\tilde{\sigma}\tilde{m}+\xi\tilde{\sigma}V & u \xi p_x +ivp_y & \tilde{g}_{0,+} + \tilde{g}_{1,+} & g_{0,-} + g_{1,-}\\
u\xi p_x -ivp_y & -\xi\tilde{\sigma}\tilde{m}+\xi\tilde{\sigma}V & g_{0,-} - g_{1,-} & \tilde{g}_{0,+} - \tilde{g}_{1,+}\\
\tilde{g}_{0,+} + \tilde{g}_{1,+} & g_{0,-} - g_{1,-} & \xi\tilde{\sigma}\tilde{m} + \xi\tilde{\sigma}V  & u\xi p_x -ivp_y\\
g_{0,-} + g_{1,-} & \tilde{g}_{0,+} - \tilde{g}_{1,+} & u\xi p_x +ivp_y &  -\xi\tilde{\sigma}\tilde{m} +  \xi\tilde{\sigma}V
\end{pmatrix}.   
 \end{eqnarray}
Clearly, the effective Hamiltonian in Eq.~\eqref{eq:CoupledMassiveDirac} describes a pair of asymmetrically coupled massive Dirac fermions with a spin-valley-dependent mass and spin-valley-dependent on-site potential $\tilde{\sigma}V$ -- a key result we use in the main text to study polarization in T$_{\rm d}$-WTe$_2$ bilayers. 

We note that, following the same argument we outlined for AB honeycomb bilayer in the main text, the effective massive Dirac model in Eq.\ \eqref{eq:CoupledMassiveDirac} can be mapped (in the vicinity of $Q, Q'$ with $\bm{p} = \bm{0}$) to a valley-dependent spinful SSH chain in the $N_z \rightarrow \infty$ limit, 
\begin{equation}\label{eq:spinfulSSH}
 H_{{\rm SSH},\xi\sigma}(k) = \sum_{\alpha=0,x,y,z}  d_{\alpha, \xi\sigma}(k) \sigma_{\alpha},
\end{equation}
where $d_{0, \xi\sigma} = \xi\sigma V$, $d_x(k) = 2g_{0,-}\cos(ka), d_y(k) = -2g_{1,-}\sin(ka)$, $d_{z, \xi\sigma}(k) = \xi\sigma m + 2 g_{1,+} \cos(ka)$. Here, we drop the ``tilde" sign in the parameters in Eq.\ \eqref{eq:CoupledMassiveDirac} in keeping with the convention employed in the main text. Given $d_{\alpha, \xi\sigma} \neq 0$ for all $\alpha = 0,x,y,z$, the solid angle subtended by the four-component $\bm{d}_{\xi\sigma}$ is expected to be nonzero (Fig.\ 1b of the main text) and thus give rise to a nonzero Berry phase in each spin sector. For illustration, we consider the limit $|g_{0,-}| = |g_{1,-}| = g$ which allows us to obtain an analytic expression for the Berry phase in the valence band for a given spin $\sigma = \pm$ and valley $\xi = \pm$:
\begin{eqnarray}\label{eq:SpinBerryPhaseV}
\gamma_{v, \xi\sigma} &=& \int_{-\pi}^{\pi} dk \left[\frac{1}{2} - \frac{\xi\sigma m + 2 g_{1,+} \cos(ka)}{2\sqrt{[\xi\sigma m + 2 g_{1,+} \cos(ka)]^2 + 4 g^2}} \right]\\\nonumber
                   &\simeq& - \frac{\xi\sigma}{4} \int_{-\pi}^{\pi} dk \frac{4g^2}{m^2 + 4 \xi\sigma m g_{1,+}\cos(ka) + 4g^2_{1,+}\cos^2(ka)} + \pi (1-\xi\sigma) \pmod{ 2\pi n, n\in \mathbb{Z}}\\\nonumber
                   &=& - \xi\sigma \frac{m g^2 \pi}{(m^2 - 4g_{1,+}^2)^{3/2} }   \pmod{ 2\pi n, n\in \mathbb{Z}}.
\end{eqnarray}
Note that given $\xi \sigma = \pm 1$, the extra $(1-\xi\sigma)\pi$ term yields an extra $0$ or $2\pi$ phase that can be incorporated in the phase ambiguity of integer multiples $2\pi n$. In reaching the second line of the integral above, we made use of the binomial expansion: $(1+x)^{-1/2} \simeq 1 - \frac{x}{2} + o(x^2)$ under the limit $m \gg g$.

Since the family of T$_{\rm d}$-bilayer TMD such as MoTe$_2$ and WTe$_2$ are semi-metallic in general, with the valence bands near $Q, Q'$ being fully filled and the conduction bands partially filled as shown in Fig.2c of the main text. Therefore, $\gamma_{c, \xi\sigma}$ from conduction bands must also be considered to account for the total polarization. Note that $\gamma_{c, \xi\sigma}$ is evaluated using Eq.2 of the main text by replacing the eigenstate $\ket{u_{-}(k)}$ of the lower energy band of the SSH chain by the eigenstate $\ket{u_{+}(k)}$ of the upper band. It is straightforward to show that this amounts to replacing the ``$-$" sign in the integrand in the first line of Eq.~\eqref{eq:SpinBerryPhaseV} to ``$+$", which leads to 
\begin{eqnarray}\label{eq:SpinBerryPhaseC}
\gamma_{c, \xi\sigma} = -\gamma_{v, \xi\sigma} \pmod{ 2\pi n, n\in \mathbb{Z}} = \xi\sigma \frac{m g^2 \pi}{(m^2 - 4g_{1,+}^2)^{3/2} }   \pmod{ 2\pi n, n\in \mathbb{Z}}.
\end{eqnarray}
The expression for $\gamma_{c, \xi\sigma}$ above is presented in Eq.5 of the main text with the band index $c$  omitted for simplicity. 

Given that all spin subbands in the valence band states near $Q,Q'$ are fully filled (see Fig.2c of the main text), contributions from $\gamma_{v, \xi\sigma}$ would eventually cancel out due to the $\sigma$-dependent sign in $\gamma_{v, \xi\sigma}$. On the other hand, the spin subbands with $\sigma= +$ and $\sigma= -$ in the conduction bands have different occupation numbers in general due to the finite spin splitting $V$ (see Fig.2c of the main text) and the contributions from $\gamma_{\xi,+}$ and $\gamma_{\xi,-}$ do not cancel each other completely. As we mention in the main text and further explain in subsection II-2 below, the polarization in T$_{\rm d}$ bilayer TMD can indeed be understood as a result of partial cancellation between $\gamma_{c,\xi\sigma}$ from different spin subbands in the conduction band states.

\section*{II. Relation between Berry phase in periodic SSH and origin of $P_z$ in bilayer SFE} 

In the main text, we point out that the origin of $P_z$ in bilayer SFEs can be identified as the Berry phases (Eq.3 and Eq.5 of the main text) obtained for a periodic SSH chain with broken intra-cell bonds in the large $N_z$ limit. In this section, we present a detailed justification for this important connection. 

The rest of the section is organized as follows: in subsection 1 below, we introduce the concept of bare polarization $P_{z,0}$ generated purely by stacking geometry. In subsection 2, we establish the Berry phase in Eq.3 and Eq.5 obtained for a periodic SSH as the primary source of $P_{z,0}$ generated by stacking geometry in bilayer SFEs in the following two steps: (i) we rigorously prove that $P_{z,0}(\bm{k} \simeq \pm \bm{K})$ near $\pm K$ points in the bilayer limit is given \textit{precisely} by the Berry phase in Eq.3 of the main text; we further demonstrate numerically that the Berry phase in Eq.5 of the main text accurately captures the $P_{z,0}(\bm{k} \simeq \pm \bm{Q})$ near $\pm Q$ points; (ii) we explicitly demonstrate that the $\bm{k}$-resolved $P_{z,0}$ in bilayer SFEs is dominated by contributions $P_{z,0}(\bm{k} \simeq \pm \bm{K})$ near $\pm K$ points for AB-stacked honeycomb bilayers and 3R-bilayer TMDs, and similarly $P_{z,0}(\bm{k} \simeq \pm \bm{Q})$ dominates the bare $P_{z,0}$ for T$_{\rm d}$-bilayer TMDs. Finally, in subsection 3 we discuss corrections from surface charge effects.\\

\subsection*{1. Bare polarization $P_{z,0}$ generated by stacking geometry}\label{ssec:2-1}

The general physical mechanism behind the formation of stacking polarization can be described as follows: we start with two identical but decoupled monolayers, with monolayer Hamiltonians $H_{1}$ and $H_{2}$ respectively. Stacking one layer on another according to the stacking geometry introduces a symmetry-breaking inter-layer coupling Hamiltonian $H_{T}$ with its form constrained by the specific stacking geometry, which leads to asymmetric distribution of probability density across the two layers, \textit{i.e.}, nonzero layer polarization in the electronic wave functions. Notably, for non-periodic bilayer systems the quantum mechanical operator $\hat{z}$ for $z$-coordinates is indeed well-defined. For convenience, in the following we set the horizontal plane which bisects the inter-layer spacing $d$ as the $z=0$ plane, and the expectation value $z_n \equiv \braket{\Psi_n|\hat{z}|\Psi_n} \neq 0$ in general for an eigenstate $\ket{\Psi_n}$ of the total bilayer Hamiltonian due to the broken symmetry introduced by $H_{T}$. Given that the stacking geometry preserves in-plane lattice translation symmetry for bilayer SFEs, the original lattice Hamiltonian, as exemplified by Eq.~\eqref{eq:tb_sic}, Eq.~\eqref{eq:H3R} and Eq.~\eqref{eq:tb_tmd}, take the general form of:
\begin{eqnarray}\label{eq:HOriginal}
H_0 &=& \sum_{\bm{k}} \Psi^{\dagger}_{\bm{k}} H_0 (\bm{k}) \Psi_{\bm{k}}, \\\nonumber
\text{where } H_0 (\bm{k}) &=& 
\begin{pmatrix}
H_{1}(\bm{k}) & H_{T}(\bm{k})\\
H^{\dagger}_{T}(\bm{k}) & H_{2}(\bm{k})
\end{pmatrix}.
\end{eqnarray}
Here, $\bm{k} = (k_x, k_y)$ denotes the in-plane Bloch momentum, $\Psi^{\dagger}_{\bm{k}} = (\Psi^{\dagger}_{1,\bm{k}}, \Psi^{\dagger}_{2,\bm{k}})$ denotes the fermionic operators with subscript $l=1,2$ labeling the layer index. Thus, the out-of-plane polarization of the bilayer system described by $H_0$ is simply given by:
\begin{eqnarray}\label{eq:defPz}
P_{z, 0} = -\frac{e}{\mathcal{V}} \sum_{n \in \text{filled}, \bm{k}} \braket{\Psi^{0}_{n \bm{k}}|\hat{z}|\Psi^{0}_{n \bm{k}}},    
\end{eqnarray}
where $\mathcal{V}$ is the total volume of the system and $\ket{\Psi^{0}_{n \bm{k}}}$ denotes the Bloch state at $\bm{k}$ of a filled band $n$ with energy $E^{0}_n(\bm{k})$, where $E^{0}_n(\bm{k})$, $\ket{\Psi^{0}_{n \bm{k}}}$ are the eigenvalues and eigenstates of $H_0(\bm{k})$. The inner product is given by the integral: $\braket{\Psi^{0}_{n \bm{k}}|\hat{z}|\Psi^{0}_{n \bm{k}}} \equiv \int d^3\bm{r} z |\Psi^{0}_{n \bm{k}}(\bm{r})|^2$. It is worth noting that $-e \braket{\Psi^{0}_{n \bm{k}}|\hat{z}|\Psi^{0}_{n \bm{k}}}$ simply describes the out-of-plane dipole moment of the Bloch electron at $\bm{k}$ in a filled electron band $n$, and Eq.~\eqref{eq:defPz} recovers the physical meaning of polarization as the volume density of dipole moments.

As a consistency check, in the simplest case where the band $n$ and its associated Bloch eigenstates $\ket{\Psi^{0}_{n \bm{k}}}$ are represented by a \textit{single} Wannier function $W^{0}_n(\bm{r})$ such that $\Psi^{0}_{n \bm{k}}(\bm{r}) = \frac{1}{\sqrt{N}} \sum_{\bm{R}} e^{i \bm{k}\cdot\bm{R}} W^{0}_n(\bm{r} - \bm{R})$, Eq.~\eqref{eq:defPz} reduces to the familiar expression of polarization in terms of $W^{0}_n(\bm{r})$~\cite{VanderbuiltS, RestaS}:
\begin{eqnarray}
P_{z,0} = -\frac{e}{\Omega} \sum_{n \in {\rm filled}} \int d^3\bm{r}  z |W^{0}_n(\bm{r})|^2
\end{eqnarray} 
where $\Omega$ is the volume of the unit cell. In the case of bilayer SFEs discussed in the main text, however, $\ket{\Psi^{0}_{n \bm{k}}}$ \textit{\textbf{cannot}} be represented by a single Wannier function --- due to the multiple orbitals involved in the lattice Hamiltonians (Eq.~\eqref{eq:tb_sic}, Eq.~\eqref{eq:H3R} and Eq.~\eqref{eq:tb_tmd}) and their $\bm{k}$-dependent inter-layer coupling $H_{T}(\bm{k})$, the Bloch states $\ket{\Psi^{0}_{n \bm{k}}}$ at different $\bm{k}$ have distinct orbital characters in general. Thus, we must use the more general expression (Eq.~\eqref{eq:defPz}) to calculate the polarization for bilayer SFEs. In the Dirac bra-ket notation, the Bloch states of a band $n$ are generally written as: 
\begin{eqnarray}
\ket{\Psi^{0}_{n \bm{k}}} &=& \sum_{l, \alpha, \sigma} c^{0}_{n,l\alpha\sigma}(\bm{k}) \ket{\Psi_{l\alpha\sigma \bm{k}}}, \\\nonumber
\text{where} \ket{\Psi_{l\alpha\sigma \bm{k}}} &=& \frac{1}{\sqrt{N}} \sum_{\bm{R}} e^{i \bm{k}\cdot\bm{R}} \sum_{l \alpha \sigma} \ket{\phi_{l\alpha\sigma}(\bm{R})}.
\end{eqnarray}
Here, the coefficients $c^{0}_{n,l\alpha\sigma}(\bm{k})$ can be obtained by exact diagonalization of the full lattice Hamiltonian $H_{0}(\bm{k})$. $\ket{\Psi_{l\alpha\sigma \bm{k}}}$ is the original Bloch basis (linear combination of atomic orbitals) in the decoupled layer $l$ with orbital/sublattice index $\alpha$ and spin index $\sigma$, which forms the basis states of the lattice Hamiltonians in Eq.~\eqref{eq:tb_sic}, Eq.~\eqref{eq:H3R} and Eq.~\eqref{eq:tb_tmd}. Note that $\bm{R}$ denotes the 2D in-plane Bravais lattice vectors, and the atomic orbital $\ket{\phi_{l\alpha\sigma}(\bm{R})}$ in layer $l$ is localized at the atomic site $(\bm{R},  R^{(l)}_z)$ with $R^{(l)}_z = \braket{\phi_{l\alpha\sigma}(\bm{R})|\hat{z}|\phi_{l\alpha\sigma}(\bm{R})} = (-1)^{l+1} \frac{d}{2}$. 

Making use of the symmetry properties of $p$- or $d$-orbitals in the lattice Hamiltonians Eq.~\eqref{eq:tb_sic}, Eq.~\eqref{eq:H3R} and Eq.~\eqref{eq:tb_tmd} and the fact that the orbitals $\ket{\phi_{l\alpha\sigma}(\bm{R})}$ are exponentially localized, we find that $\braket{\phi_{\sigma'\alpha'} ( \bm{R}', R^{(l')}_z) |z | \phi_{\sigma\alpha} (\bm{R}, R^{(l)}_z)} \simeq R^{(l)}_z \delta_{\bm{R} \bm{R}'}\delta_{ll'} \delta_{\sigma \sigma'} \delta_{\alpha \alpha'}$ holds in general. This helps to simplify Eq.~\eqref{eq:defPz} as
\begin{eqnarray}\label{eq:sumPk}
P_{z,0} &=& \frac{1}{N} \sum_{\bm{k} \in BZ} P_{z,0} (\bm{k}) \\\nonumber
\text{with } P_{z,0} (\bm{k}) &=& \frac{-e}{\Omega} \sum_{n \in \text{filled}} \sum_{l \sigma \alpha} |c^{0}_{n, l\sigma \alpha} (\bm{k})|^2 R^{(l)}_z \\\nonumber
&=& P_0 \left\{ \frac{1}{2} \left[ \sum_{n \in \text{filled}} \big( \sum_{\sigma \alpha} |c^{0}_{n, 1\sigma \alpha} (\bm{k})|^2 - \sum_{\sigma \alpha}|c^{0}_{n, 2 \sigma \alpha} (\bm{k})|^2 \big)\right] \right\} \\\nonumber
&=& \frac{P_0}{2} \left[ \sum_{n \in \text{filled}} \big( |\Psi_{1, n} (\bm{k})|^2 - |\Psi_{2, n} (\bm{k})|^2 \big) \right].
\end{eqnarray}
Here, $P_0 \equiv -e/\Omega_0$ is the polarization quantum with $\Omega_0$ denoting the area of the 2D unit cell (note: $\Omega_0 d = \Omega$). $|\Psi_{l, n} (\bm{k})|^2$ represents the number density at $\bm{k}$ of band $n$ in layer $l$. Thus, Eq.~\eqref{eq:sumPk} essentially recovers the physical meaning of polarization as the number density difference between two different layers. We note that similar formula as Eq.~\eqref{eq:sumPk}, \textit{i.e.}, summation over $\bm{k}$-resolved number density difference across the layers, have been adopted to account for out-of-plane polarization in bilayer systems in the literature, see for example, Eq.S8 in the Supplementary Information of Refs.~\cite{ZhengS}, and Eq.2 in Ref.~\cite{HongyiS}.\\

\subsection*{2. Berry phase origin of bare polarization $P_{z,0}$ in bilayer SFE}\label{ssec:2-2}

The expression of $P_{z,0}$ in Eq.~\eqref{eq:sumPk} allows us to resolve the polarization in 2D $\bm{k}$-space perpendicular to the polarization in $z$-direction, with the component $P_{z,0}(\bm{k})$ being $\bm{k}$-dependent in general. Next, we discuss the relation between $P_{z,0}(\bm{k})$ at $K$, $Q$ points to the Berry phases in Eq.3 and Eq.5 of the main text for periodic SSH chains.

\subsubsection*{A. Relation between $P_{z,0}$ near $\pm K$ points and Berry phase (Eq.3) in periodic SSH chain}

We first consider honeycomb bilayers and 3R-bilayer TMDs, where the relevant effective Hamiltonians at $\pm K$ are mapped to the two-cell limit of an SSH chain (see Eq.~\eqref{eq:tb_sic_k} and Eq.~\eqref{eq:Heff_mos2}). We note that it suffices to consider $\bm{k} = +\bm{K}$ as the polarization at $-\bm{K}$ is given by $P_{z,0}(\bm{-\bm{K}}) = P_{z,0}(\bm{+\bm{K}})$ following time-reversal symmetry. With $\epsilon_B < \epsilon_A$ and $\Delta_{AB} \equiv \frac{\epsilon_A - \epsilon_B}{2} \gg g_{BA}$ (see Table~\ref{tab:sic1}), the valence bands states at $\bm{K}$ consist of two lower-lying eigestates of the 4$\times$4 Hamiltonian $H_{AB}(\bm{K})$ in Eq.~\eqref{eq:tb_sic_k}: 
\begin{eqnarray}
(\textrm{i}) \ket{v,1, \bm{K}} &=& c_{v,1B}(\bm{K}) \ket{1,B} + c_{v,2A}(\bm{K}) \ket{2,A}, \text{with energy }  E_{v,1}(\bm{K}) = \frac{\epsilon_A + \epsilon_B}{2} - \sqrt{\Delta^2_{AB} + g^2_{BA}},  \\\nonumber
\text{where }  c_{v,1B}(\bm{K}) &=& \frac{\Delta_{+}}{\sqrt{\Delta^2_{+} + g^2_{BA}}}, c_{v,2A}(\bm{K}) = \frac{-g_{BA}}{\sqrt{\Delta^2_{+} + g^2_{BA}}} \text{ with } \Delta_{+} \equiv \Delta_{AB} + \sqrt{\Delta^2_{AB} + g^2_{BA}}. \\\nonumber
(\textrm{ii}) \ket{v,2, \bm{K}} &=& c_{v,2B}(\bm{K}) \ket{2,B}, \text{ with energy } E_{v,2}(\bm{K}) = \epsilon_B \text{ and } c_{v,2B}(\bm{K}) = 1.
\end{eqnarray}\\
The value of $P_{z,0}(\bm{K})$ according to Eq.~\eqref{eq:sumPk} is then given by:
\begin{eqnarray}\label{eq:PKBerryPhase}
P_{z,0}(\bm{K}) &=& \frac{P_0}{2} \big[ |c_{v,1B}(\bm{K})|^2 - |c_{v,2A}(\bm{K})|^2 - |c_{v,2B}(\bm{K})|^2 \big]\\\nonumber
&=& -\frac{P_0}{2} \big[ 1 + \frac{g^2_{BA}}{\Delta^2_{+} + g^2_{BA}} - \frac{\Delta^2_{+}}{\Delta^2_{+} + g^2_{BA}} \big] \\\nonumber
&=& -\frac{P_0}{2} \big[ 1 - \frac{2\Delta^2_{AB} + 2\Delta_{AB}\sqrt{\Delta^2_{AB} + g^2_{BA}}}{2(\Delta^2_{AB} + g^2_{BA}) +2\Delta_{AB}\sqrt{\Delta^2_{AB} + g^2_{BA}}} \big] \\\nonumber
&=&-\frac{P_0}{2} \big[ 1 - \frac{\Delta_{AB} (\Delta_{AB}+\sqrt{\Delta^2_{AB} + g^2_{BA}})}{\sqrt{\Delta^2_{AB} + g^2_{BA}}(\Delta_{AB} + \sqrt{\Delta^2_{AB} + g^2_{BA}})} \big] \\\nonumber
&=&-\frac{P_0}{2} \big[ 1 - \frac{\Delta_{AB}}{\sqrt{\Delta^2_{AB} + g^2_{BA}}} \big].
\end{eqnarray}

By identifying $g_{BA} \equiv 2t$ where $2t$ stands for the inter-cell hopping in the 1D SSH chain as defined in the main text, we find that $P_{z,0}(\bm{K}) = -\frac{P_0}{2\pi} \gamma$, where $\gamma$ is \textit{exactly} the Berry phase in Eq.3 obtained for a periodic SSH chain. This unambiguously verifies that $P_{z,0}(\bm{k} \simeq \pm\bm{K})$ in AB-stacked honeycomb bilayers and 3R-MoS$_2$ originates exactly from the Berry phase in Eq.3 of the main text.

\subsubsection*{B. Relation between $P_{z,0}$ near $\pm Q$ points and Berry phase (Eq.5) in periodic spinful SSH chain}

Now we discuss the case of T$_{\rm d}$ bilayer TMD family which are metallic in general with electron bands near $Q,Q'$ being partially filled (see Fig.2c of the main text). As we mention in the discussions on Eq.~\eqref{eq:SpinBerryPhaseV}-Eq.~\eqref{eq:SpinBerryPhaseC} at the end of Section I, the polarization of T$_{\rm d}$ bilayer TMD should be understood as a result of partial cancellation between $\gamma_{c, \xi\sigma}$ in the partially filled conduction bands. Here we provide a detailed justification.

Note that even for a metallic system where some electronic bands are partially filled, the polarization would be given by summing over contributions from all filled Bloch states, and Eq.~\eqref{eq:defPz} is generalized as:
\begin{eqnarray}\label{eq:defPzmetal}
P_{z, 0} = -\frac{e}{\mathcal{V}} \sum_{n, \bm{k}} \braket{\Psi^{0}_{n \bm{k}}|\hat{z}|\Psi^{0}_{n \bm{k}}} f_{n \bm{k}}, 
\end{eqnarray}
where $f_{n \bm{k}}$ is the occupation number of the Bloch state at momentum $\bm{k}$ in band $n$, with $f_{n\bm{k}} = 1$ for $E_{n\bm{k}} \leq \mu$ and $f_{n\bm{k}} = 0$ for $E_{n\bm{k}} > \mu$. 

To see how $P_{z,0}$ is related to $\gamma_{c, \xi\sigma}$ in Eq.5 (Eq.~\eqref{eq:SpinBerryPhaseC}), we note that for each in-plane momentum $\bm{k}$, the polarization $P_{z,0}(\bm{k})$ at $\bm{k}$ is related to the sum of Berry phases $\gamma_{n, \bm{k}}$ over all filled states below the Fermi level at $\bm{k}$ by: 
\begin{equation}
P_{z,0}(\bm{k}) = \frac{P_0}{2\pi} \sum_{n} \gamma_{n, \bm{k}} f_{n\bm{k}},
\end{equation}
where $n$ runs over all bands and the occupation function $f_{n\bm{k}}$ automatically excludes contributions from unfilled states with $E_{n\bm{k}} > \mu$. Thus, according to $P_{z,0} = \sum_{\bm{k}} P_{z,0}(\bm{k})/N$ 
 (see Eq.~\eqref{eq:sumPk}) we can rewrite $P_{z,0}$ as: 
\begin{equation}
P_{z,0} = \frac{P_0}{2\pi N} \sum_{n, \bm{k}} \gamma_{n, \bm{k}} f_{n\bm{k}}.
\end{equation}

As we shall confirm numerically in subsection C below, $P_{z,0}$ for bilayer T$_{\rm d}$-TMD has a dominant contribution near $Q, Q'$ with $\bm{k} \simeq \xi \bm{Q}$ ($\xi$: valley index for $Q, Q'$). This motivates us to focus only on momenta $\bm{k} \simeq \xi \bm{Q}$ within a neighborhood of $\xi\bm{Q}$ in the summation above, where we have $\gamma_{n} (\bm{k} = \bm{p}+ \xi \bm{Q}) \simeq \gamma_{n}(\xi\bm{Q})$. Note that $\gamma_{n}(\xi\bm{Q})$ involves $\gamma_{v, \xi \sigma}$ (Eq.~\eqref{eq:SpinBerryPhaseV}) for the two spin subbands in valence bands: $n = (v, +), (v, -) $, as well as $\gamma_{c, \xi \sigma}$ (Eq.~\eqref{eq:SpinBerryPhaseC}) for the two spin subbands in conduction bands: $n = (c, +), (c, -) $. Thus, the total $P_{z,0}$ can be approximately written as:
\begin{eqnarray}\label{eq:PzImbalance}
P_{z,0} &\simeq& \frac{P_0}{2\pi N} \big[\sum_{\xi \sigma, \bm{k}\sim\bm{Q}} \gamma_{c, \xi \sigma} f_{c, \xi\sigma}(\bm{k}) + \sum_{\xi \sigma, \bm{k}\sim\bm{Q}} \gamma_{v, \xi \sigma} f_{v, \xi\sigma}(\bm{k}) \big]\\\nonumber
        &=&  \frac{P_0}{2\pi} \big[\sum_{\xi\sigma} \gamma_{c, \xi\sigma} \frac{\sum_{\bm{k}\sim\bm{Q}} f_{c, \xi\sigma}(\bm{k})}{N} + \sum_{\xi\sigma} \gamma_{v, \xi\sigma} \frac{\sum_{\bm{k}\sim\bm{Q}} f_{v, \xi\sigma}(\bm{k})}{N} \big] \\\nonumber
        &=&  \frac{P_0}{2\pi} \sum_{\xi\sigma} \gamma_{c, \xi\sigma} \nu_{c, \xi\sigma} \\\nonumber
       &=&  2 \frac{P_0}{2\pi} (\gamma_{c, ++} \nu_{c, ++} + \gamma_{c, +-} \nu_{c, +-} )  \\\nonumber
(\gamma_{c, ++} = - \gamma_{c, +-} \equiv + \gamma_c)       &=& 2 \frac{P_0}{2\pi}  (\nu_{c, ++} - \nu_{c, +-} ) \gamma_c,
\end{eqnarray}
where $\nu_{c, \xi\sigma} \equiv \frac{\sum_{\bm{k}\sim\bm{Q}} f_{c, \xi\sigma}(\bm{k})}{N} $ ($\nu_{v, \xi\sigma} \equiv \frac{\sum_{\bm{k}\sim\bm{Q}} f_{v, \xi\sigma}(\bm{k})}{N}$) denotes the filling factors of conduction (valence) band with spin $\sigma = \pm$ at valley $\xi = \pm$, and $\gamma_c = |\gamma_{c, \xi\sigma}|$ is the amplitude of the Berry phase in Eq.5 of the main text (Eq.~\eqref{eq:SpinBerryPhaseC}). The extra factor of 2 in the fourth and fifth lines accounts for the equal contributions from both valleys $\xi = +$ and $\xi = -$, which follows from the relations $\gamma_{c, \xi, \sigma} = \gamma_{c, -\xi, -\sigma}$ and $\nu_{c, \xi, \sigma} = \nu_{c, -\xi, -\sigma}$ imposed by time-reversal symmetry $\mathcal{T}$.

Note that in the third line above we drop contributions from the valence bands because both valence bands with $\sigma= +$ and $\sigma= -$ are fully filled near $Q$ and $Q'$(Fig.2c of the main text) which implies $f_{v,\xi, +}(\bm{k}) = f_{v,\xi, -}(\bm{k}) = 1$ for $\bm{k} \sim \xi \bm{Q}$, and contributions from opposite spins $\sigma = \pm$ cancel out each other given $\gamma_{v,\xi, +} = - \gamma_{v,\xi, -}$. On the other hand, the conduction bands near $Q$ and $Q'$ are partially filled, and the valley-dependent spin splitting $\xi\sigma V$ causes $\nu_{c, \xi,+} \neq \nu_{c, \xi, -}$ in general (see Fig.2c of main text), which leads to a finite total Berry phase summed over all filled states and results in $P_{z,0} \neq 0$. 

To confirm that the origin of $P_{z,0}$ can indeed be traced down to $\gamma_{c}$ according to Eq.~\eqref{eq:PzImbalance}, we obtain $\gamma_c \simeq 0.03 \pi$ by setting $g \equiv |g_{1,-}| = 0.03$ eV, $g_{1,+} = 0.03$ eV and $m = 0.09$ eV in Eq.5 (Eq.~\eqref{eq:SpinBerryPhaseC}) relevant for bilayer WTe$_2$, and $\nu_{c,++} - \nu_{c,+-} \approx 2.6 \times 10^{-3}$ by numerically counting the filled conduction band states for spin $\sigma=\pm$ below $\mu = 0$. Given the polarization quantum $P_0 = e/(ab) = 72.655$ $\mu$C/cm$^2$ with $a = 3.49 {\AA}$ and $b = 6.31 {\AA}$ for WTe$_2$, Eq.~\eqref{eq:PzImbalance} gives $P_{z,0} \simeq (\nu_{c,++} - \nu_{c,+-}) \gamma_c P_0/\pi = 0.057$ $\mu$C/cm$^2$. On the other hand, using the full lattice Hamiltonian (Eq.~\eqref{eq:tb_tmd}) for bilayer WTe$_2$ with $\mu = 0$ and all $\bm{k}$ points included, we obtain $P_{z,0} = 0.0617$ $\mu$C/cm$^{2}$ according to Eq.~\eqref{eq:defPzmetal}. The reasonably good agreement with realistic lattice model calculations justifies that the relation between $\gamma_{c, \xi\sigma}$ in Eq.5 of the main text to the bare stacking polarization in Eq.~\eqref{eq:PzImbalance} has well captured the origin of $P_{z,0}$ in bilayer WTe$_2$. The analysis above also explains why the polarity of bilayer T$_{\rm d}$-TMD is much weaker than honeycomb bilayer and 3R-TMDs: the total Berry phase results from a small imbalance between two spin subbands with opposite Berry phases.

\subsubsection*{C. Dominant contributions to $P_{z,0}$ near $K$ and $Q$ points}

Having established the relations between the Berry phases in Eq.3 and Eq.5 of the main text to the bare polarization $P_{z,0}$ at $\pm K$ and $Q,Q'$ points in bilayer SFEs, we now explicitly demonstrate that the total $P_{z,0}$ is dominated by contributions from $\pm K$ points in bilayer SiC and 3R-MoS$_2$, and those from $Q, Q'$ points in bilayer T$_{\rm d}$ WTe$_2$. This leads us to conclude that the Berry phases in Eq.3 and Eq.5 provide the primary source of the total $P_{z,0}$ in bilayer SFEs.\\

\noindent\textbf{Type I: AB-stacked honeycomb bilayers (SiC)}\\
  The area of the 2D unit cell is given by $\Omega_0=\frac{\sqrt{3}a^2}{2}$ with lattice constant $a=3$ {\AA}. The value of bare polarization is found to be $P_{z,0}=1.84~\mu$C/cm$^2$ via Eq.~\eqref{eq:sumPk} with $\ket{\Psi^{0}_{n\bm{k}}}$ solved by exact diagonalization of $H_{AB}(\bm{k})$ in Eq.~\eqref{eq:tb_sic}. The tight-binding parameters used for $H_{AB}(\bm{k})$ are tabulated in Table~\ref{tab:sic1}. The $\bm{k}$-resolved bare polarization $P_{z,0}(\bm{k})$ is shown in Fig.~\ref{fig:contour_supp}(a) where we observe dominant contributions from $K$ and $K'$ valleys. In particular, to numerically verify the dominance of contributions near $\pm K$, we calculate the contributions to $P_{z,0}$ within a neighborhood of $\pm K$ with a momentum cutoff $k_c \simeq 0.53$ {\AA}$^{-1}$ measured from $\pm K$ points (regions enclosed by dashed blue circles in Fig.~\ref{fig:contour_supp}(a)), which we denote as $P'_{z,0} \equiv \frac{2}{N}\sum_{\bm{k}: |\bm{k} -  \bm{K}|<k_c} P_{z,0}(\bm{k})$ (the factor of 2 accounts for the equal contributions from the two inequivalent $\pm K$ valleys due to time-reversal symmetry:$P_{z,0}(\bm{k}) = P_{z,0}(-\bm{k})$). The value of $k_c$ is set by momentum points where half of the maximum value $P_{z,0}(\pm \bm{K})$ at $\pm K$ is attained. We find that contributions from the neighborhood of $\pm K$ take up $p_{K} = P'_{z,0}/P_{z,0} \approx 71 \%$ as tabulated in Table~\ref{table:PzwithRatio}.\\

\noindent\textbf{Type II: Rhombohedral (3R) bilayer TMDs (3R-MoS$\bm{_2}$)}\\
The area of the 2D unit cell is given by $\Omega_0=\frac{\sqrt{3}a^2}{2}$ with lattice constant $a=3.2$ {\AA}. The value of bare polarization is found to be $P_{z,0}=-0.653~\mu$C/cm$^2$ via Eq.~\eqref{eq:sumPk} with $\ket{\Psi^{0}_{n\bm{k}}}$ solved by exact diagonalization of $H_{3R}(\bm{k})$ in Eq.~\eqref{eq:H3R}. The tight-binding parameters used are tabulated in Table~\ref{table:MoS2}. The $\bm{k}$-resolved bare polarization $P_{z,0}(\bm{k})$ is shown in Fig.~\ref{fig:contour_supp}(b) where we observe dominant contributions from $K$ and $K'$ valleys. Using the same scheme for bilayer SiC discussed above, we confirm that the majority of contributions to $P_{z,0}$ originates from the neighborhood of $K$ and $K'$ valleys (regions enclosed by dashed blue circles in Fig.~\ref{fig:contour_supp}(b), where the cutoff momentum $k_c \simeq 0.5$ {\AA}$^{-1}$ measured from $\pm K$ points is obtained by momentum points $\bm{k}$ for which $P_{z,0}(\bm{k})$ reaches half of the full maximum value at $P_{z,0}(\bm{k} = \pm \bm{K})$). We numerically verify that $p_{K} = P'_{z,0}/P_{z,0} \approx 50 \%$ as tabulated in Table~\ref{table:PzwithRatio}. 

On the other hand, we note that there exist relatively minor but non-negligible contributions from regions near the time-reversal $M_1, M_2, M_3$ points (regions encircled by the yellow dashed rectangles in Fig.~\ref{fig:contour_supp}(b)), which take up approximately $30 \%$ of the total $P_{z,0}$ (Table~\ref{table:PzwithRatio}). As we discuss in detail in section III below, the effective Hamiltonians near these $M$-points can also be mapped to an SSH chain with Berry phase of the same form as Eq.3 of the main text, which accurately captures $P_{z,0}(\bm{k} \simeq \bm{M})$ in the neighborhood of the $M$-points.\\

\noindent\textbf{Type III: Bilayer T$_{\rm d}$-MX$_2$ (T$_{\rm d}$-WTe$\bm{_2}$)}\\
The area of the 2D unit cell is $\Omega_0=ab$ with lattice constants $a=3.49$ {\AA} and $b=6.31$ {\AA}. The value of bare polarization is found to be $P_{z,0}=-0.062~\mu$C/cm$^2$ via Eq.~\eqref{eq:sumPk} with $\ket{\Psi^{0}_{n\bm{k}}}$ solved by exact diagonalization of $H_{T_{\rm d}}(\bm{k})$ in Eq.~\eqref{eq:tb_tmd}.  The parameters used for WTe$_2$ are presented in Tables~\ref{tab:td_s1} and \ref{tab:td_s2}. The $\bm{k}$-resolved $P_{z,0}$ is plotted in Fig.~\ref{fig:contour_supp}(c). We observe that the polarization is concentrated at the $Q$ and $Q'$ pockets in the Brillouin zone, consistent with our analysis in Section II-2B. To numerically verify the dominance of contributions near $Q, Q'$, we calculate the contributions to $P_{z,0}$ within the region enclosed by the outer Fermi surface contour of filled conduction bands which enclose the $Q, Q'$ points (dark blue dashed lines in Fig.~\ref{fig:contour_supp}(c). We find the percentage of contributions from the neighborhood of $Q,Q'$ to be $p_{Q} = P'_{z,0}/P_{z,0} \approx 92 \%$ as tabulated in Table ~\ref{table:PzwithRatio}.\\

\subsection*{3. Surface charge corrections and self-consistency equation}\label{ssec:2-3}

\begin{figure}
    \centering
    \includegraphics[width=\linewidth]{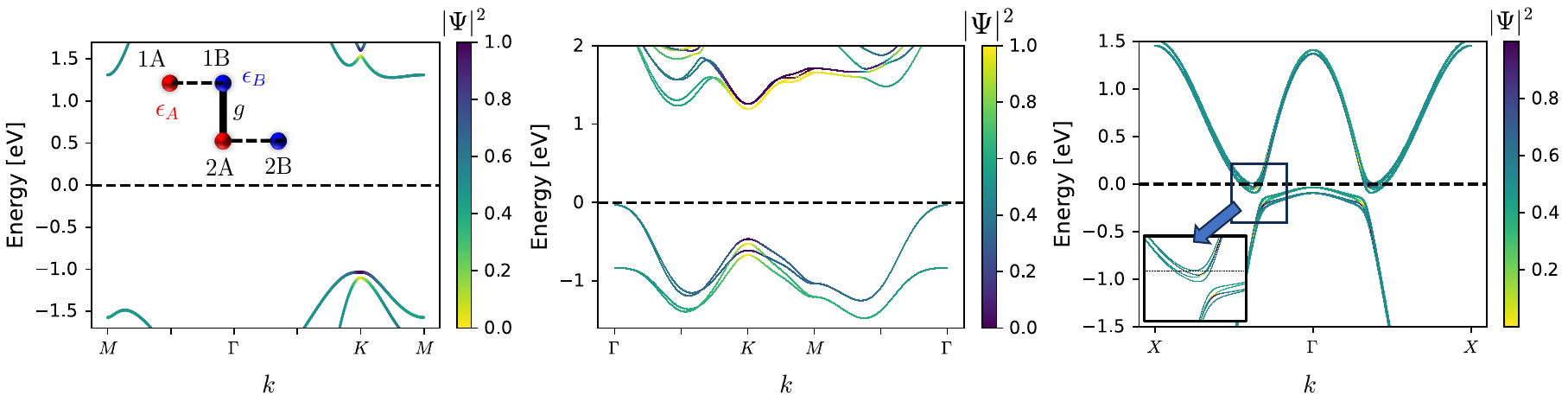}

    \caption{\textbf{(Left)} Tight-binding band structure of AB-stacked bilayer SiC with the colorbar indicating the weight from the top layer in the eigenstates. Inset: schematic of the AB stacking configuration. The band gap at $K$ is given by $\epsilon_A - \epsilon_B \approx 2.2$ eV. The band structure is obtained by the tight-binding model (parameters listed in Table~\ref{tab:sic1}), with self-consistently determined $\Phi = 0.073$ eV by solving Eq.~\eqref{eq:selfconsistenteqn}. \textbf{(Middle)} Bandstructure of 3R-MoS$_2$ using the parameters in Table~\ref{table:MoS2} with $\Phi=0.05$~eV. \textbf{(Right)} Band structure of T$_{\rm d}$-WTe$_2$ based on the tight-binding model of Eq.~\eqref{eq:tb_tmd} and self-consistently obtained $\Phi=0.0015$ eV through Eq.~\eqref{eq:selfconsistenteqn}. The tight-binding model parameters are presented in Table~\ref{tab:td_s1} and the interlayer coupling parameters in Table~\ref{tab:td_s2}. The inset shows a zoom-in plot of energy bands at $Q$.}
    \label{fig:fig_supp}
\end{figure}

Our discussions through subsections II-1 and II-2 above have so far been focusing on the bare polarization $P_{z,0}$ generated purely by stacking geometry based on the general bilayer Hamiltonian in Eq.~\eqref{eq:HOriginal}. Note that our starting point is two identical and decoupled monolayers with the same electrostatic potentials: $\Phi_1 = \Phi_2$. The emergence of a bare polarization $P_{z,0}$ under nontrivial stacking geometry, however, would induce bound charges of opposite signs on the top and bottom layers and this effect cannot be ignored in the bilayer limit as we discuss in the main text. This induces a nonzero potential difference $\Phi = \Phi_1 - \Phi_2 \neq 0$ across the two layers, and the total bilayer Hamiltonian under realistic conditions must incorporate an extra potential difference term $H_{\Phi}$ into the original bilayer Hamiltonian $H_0$ in Eq.~\eqref{eq:HOriginal}:
\begin{eqnarray}\label{eq:HReal}
H &=& \sum_{\bm{k}} \Psi^{\dagger}_{\bm{k}} H (\bm{k}) \Psi_{\bm{k}}, \\\nonumber
\text{where } H (\bm{k}) &=& H_0 (\bm{k}) + H_{\Phi} \\\nonumber
&=&
\begin{pmatrix}
H_{1}(\bm{k}) + \frac{\Phi}{2}\times I_{{\rm N} \times {\rm N}} & H_{T}(\bm{k})\\
H^{\dagger}_{T}(\bm{k}) & H_{2}(\bm{k}) - \frac{\Phi}{2}\times I_{{\rm N} \times {\rm N}}
\end{pmatrix},
\end{eqnarray}
where $\rm N$ in the identity matrix $I_{{\rm N} \times {\rm N}}$ denotes the total number of relevant degrees of freedom within each monolayer. Note that due to the extra $H_{\Phi}$-term the new Bloch states of $H$ are generally functions of $\Phi$ which we denote as $\Psi_{n \bm{k}}(\Phi)$, and they are generally different from the unperturbed $\Psi^{0}_n(\bm{k}) \equiv \Psi_{n \bm{k}}(\Phi = 0)$ of the original stacking Hamiltonian $H_0$ with $\Phi = 0$. Thus, the value of $P_{z}$ must also be reevaluated by replacing $\Psi^{0}_n(\bm{k})$ by $\Psi_n(\bm{k})(\Phi)$ in the defining equation (Eq.~\eqref{eq:defPz}):
\begin{eqnarray}\label{eq:defPzReal}
P_{z}(\Phi) = -\frac{e}{\mathcal{V}} \sum_{n \in \text{filled}, \bm{k}} \braket{\Psi_{n \bm{k}}(\Phi)|\hat{z}|\Psi_{n \bm{k}}(\Phi)}.    
\end{eqnarray}

\begin{figure*}
    \centering
    \includegraphics[width=\linewidth]{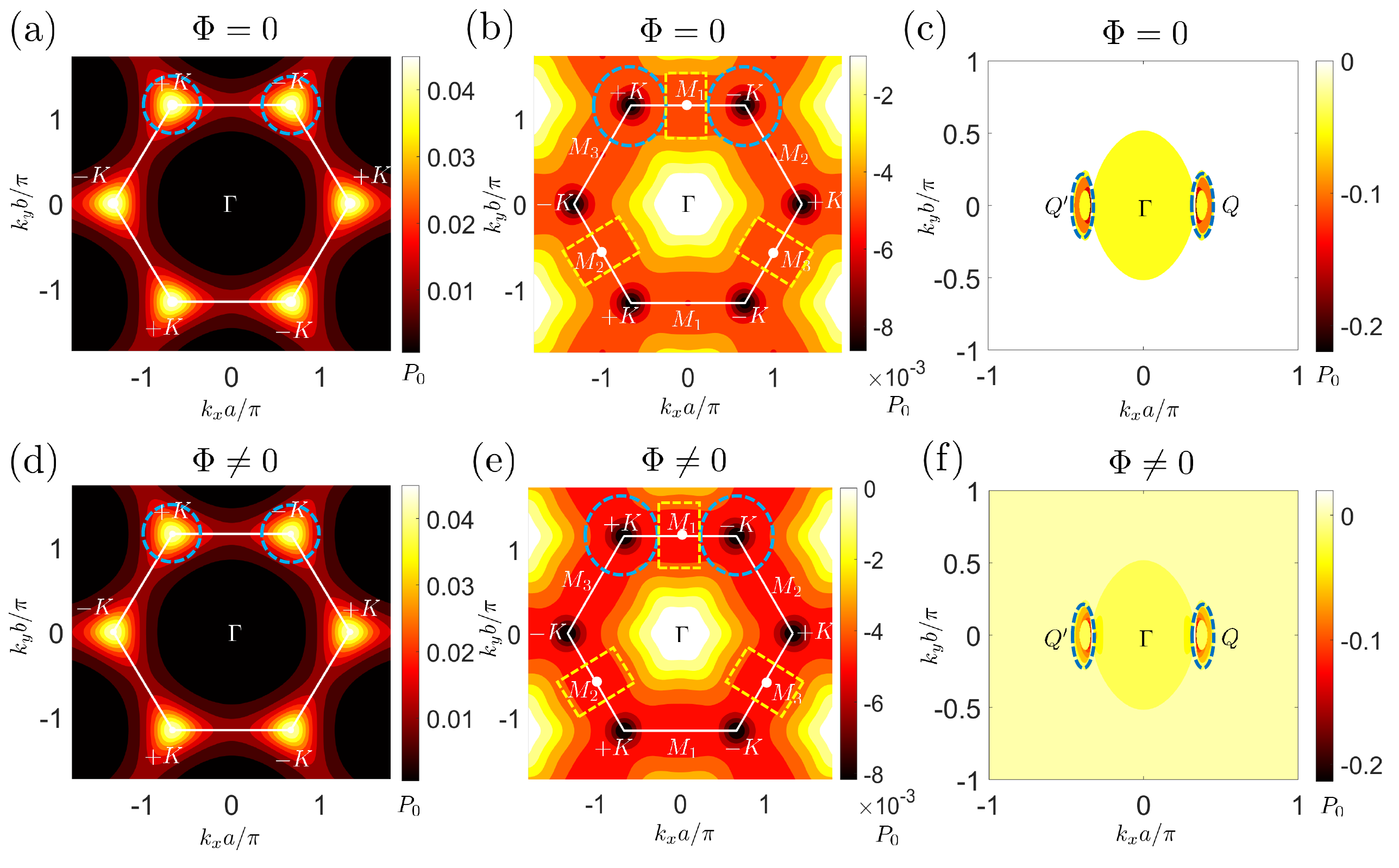}
    \caption{Contour plots of $\bm{k}$-resolved bare polarization $P_{z,0}$ obtained from Eq.~\eqref{eq:sumPk} under $\Phi = 0$ for (a) bilayer SiC, (b) bilayer 3R-MoS$_2$, and $(c)$ bilayer T$_{\rm d}$-WTe$_2$. (d)-(f): Same contour plots for eventual $P_z$ after corrections from $\Phi \neq 0$. Evidently, dominant contributions arise from regions around $\pm K$ in SiC (dashed blue circles in (a) and (d)) and 3R-MoS$_2$ (dashed blue circles in (b) and (e)), and from regions around $Q,Q'$ (encircled by dashed dark blue lines in (c) and (f)) in WTe$_2$. The percentage of contributions from each region enclosed by dashed lines are presented in Table~\ref{table:PzwithRatio}. Note that non-neligible contributions also arise near the three time-reversal invariant $M_1$, $M_2$, $M_3$ points in 3R-MoS$_2$ (dashed yellow rectangles in (b) and (e)). Detailed discussion of polarization physics near $M$-points is presented in Section III-2.}
    \label{fig:contour_supp}
\end{figure*}

On the other hand, from classical electrostatics we know that $P_{z}$ and $\Phi$ are related through the simple relation $\Phi = -\frac{e P_z d_z}{\epsilon_{r}\epsilon_0}$, where $d_z$ is the inter-layer distance, $\epsilon_r$ is the relative permittivity of the material, $\epsilon_0$ is the vaccuum permittivity. Thus, we have now two ways to obtain $P_z$ for a given $\Phi$: one through the defining relation in Eq.~\eqref{eq:defPzReal} derived from the microscopic quantum theory, while the other through $\Phi = -\frac{e P_z d_z}{\epsilon_{r}\epsilon_0}$ required by electrostatics. For the two approaches to agree with each other, for a given $\Phi$, the value of $P_z(\Phi)$ obtained from Eq.~\eqref{eq:defPzReal} must produce exactly the same $\Phi$ according to electrostatics. Therefore, $P_z$ and $\Phi$ must satisfy the following self-consistency equation:
\begin{eqnarray}\label{eq:selfconsistenteqn}
\Phi &=& \frac{e^2 d_z}{\epsilon_{r}\epsilon_0\mathcal{V}} \sum_{n \in \text{filled}, \bm{k}} \braket{\Psi_{n \bm{k}}(\Phi)|\hat{z}|\Psi_{n \bm{k}}(\Phi)}, \\\nonumber
\text{or equivalently, } P_{z} &=& -\frac{e}{\mathcal{V}} \sum_{n \in \text{filled}, \bm{k}} \braket{\Psi_{n \bm{k}}[\Phi(P_z)]|\hat{z}|\Psi_{n \bm{k}}[\Phi(P_z)]}.    
\end{eqnarray}
For the sake of simplicity we present the second line of the self-consistency equations above as Eq.6 of the main text, where the dependence of $\Phi$ on $P_z$ is assumed implicitly. We note that the physics regarding Eq.~\eqref{eq:selfconsistenteqn} can be summarized as follows: the potential difference $\Phi$ generated by the internal field due to nonzero $P_z$ according to electrostatics would in turn affect the evaluation of $P_z$ at the quantum mechanical level --- this feedback mechanism is illustrated schematically in Fig.2d of the main text. It is important to note that the physics of surface charge corrections discussed above is required by the fundamental electrostatic laws and the quantum theory of polarization, thus it applies to general bilayer ferroelectrics and does not depend on specific details of the stacking geometry or microscopic models.

Due to the correction from $\Phi \neq 0$, the eventual value of polarization for bilayer SFE is clearly different from the bare stacking polarization: $P_z \neq P_{z,0}$. However, we emphasize that the origin of the total $P_z$ should always be identified as the $P_{z,0}$ obtained from the original Hamiltonian (Eq.~\eqref{eq:HOriginal}) because $P_{z,0}$ provides the primary source of the polarization. Corrections from $\Phi \neq 0$ must be viewed only as a secondary effect based on realistic considerations for bilayer systems, and importantly, a nonzero $\Phi$ would not be present if $P_{z,0} = 0$ in the first place. Thus, the complications from surface charge corrections do not affect our conclusion that the origin of the polarization in bilayer SFEs are rooted in the nontrivial Berry phases in Eq.3 and Eq.5 of the main text, whose relations to the bare $P_{z,0}$ are unambiguously revealed through our analyses in subsections II-1 and II-2 above. 

It is important to note that because $P_{z,0}$ serves as the primary source of the stacking polarization, the sign of both $P_z$ and $\Phi$ must be governed by the sign of $P_{z,0}$: for example, if $P_{z,0} >0$, according to Eq.~\eqref{eq:sumPk} this implies the initial electron density in layer 2 is higher than that in layer 1: $n_{2,0} > n_{1,0}$. As the surface charge correction follows as a secondary effect of $P_{z,0}$, the eventual out-of-plane polarity cannot be switched, or equivalently, the electron number density difference across the layers should not be reversed. Therefore, $n_2 > n_1$ and $P_z>0$ must still hold after corrections with $\Phi = \Phi_1 - \Phi_2 <0$. On the practical level, when solving the self-consistent equations in Eq.~\eqref{eq:selfconsistenteqn}, one must seek self-consistent solutions of $P_z$ and $\Phi$ with their signs compatible with the sign of $P_{z,0}$ obtained using Eq.~\eqref{eq:defPz} or Eq.~\eqref{eq:defPzmetal} based on Bloch states of the original bilayer Hamiltonian (Eq.~\eqref{eq:HOriginal}) under $\Phi = 0$.

The eventual values of $P_z$ after correction from $\Phi \neq 0$ are presented in Table I of the main text, and the percentage of contributions from important $\bm{k}$-space regions such as $\pm K$, $Q,Q'$ and $M$-points are tabulated in Table~\ref{table:PzwithRatio}. Contour plots of $\bm{k}$-resolved $P_z$ are presented in Fig.2(e)-(g) of the main text. The $\bm{k}$-space regions near special points $\pm K$, $M_1, M_2, M_3$ and $Q, Q'$ considered in calculating the percentage of their contributions to total $P_{z,0}$ and total $P_z$ are indicated by areas enclosed by dashed lines in Fig.~\ref{fig:contour_supp}. It is evident from the contour plots in Fig.~\ref{fig:contour_supp} and Table~\ref{table:PzwithRatio} that the complications from surface charge effects do not affect the $\bm{k}$-space distribution of the polarization in a qualitative way, although the value at each $\bm{k}$, the total value of $P_z$, and the percentage of contributions from each region are subject to minor corrections from $P_{z,0}$ due to $\Phi \neq 0$. This reinforces our statement above that the surface charge corrections only serve as a secondary effect and $P_{z,0}$ provides the primary source of the polarization. Details of self-consistent calculations performed for the three types of SFEs are presented below.\\

\setlength{\tabcolsep}{6pt} 
\begin{table*}[t]
\centering 
\caption{Comparison between bare polarization $P_{z,0}$ obtained from $H_{0}(\bm{k})$ using Eq.~\eqref{eq:defPz} and Eq.~\eqref{eq:defPzmetal} with surface charge correction neglected ($\Phi=0$) and eventual polarization $P_z$ after correction from $\Phi\neq 0$ obtained by solving Eq.~\eqref{eq:selfconsistenteqn} in different types of SFE materials. Important $\bm{k}$-space regions with contributions taking up a percentage $p>30 \%$ of the total $P_{z,0}$ and total $P_z$ (enclosed by dashed lines in Fig.~\ref{fig:contour_supp}) are listed for each material.}
\begin{tabular}{c c c c}
\hline\hline
Bilayer SFE materials & AB-stacked bilayer SiC & 3R-bilayer MoS$_2$ & T$_{\rm d}$-bilayer WTe$_2$\\
\hline
$P_{z,0} (\Phi = 0)$ &  1.84 $\rm \mu C /{\rm cm}^{2}$  &  0.653 $\rm \mu C /{\rm cm}^{2}$   &  0.062 $\rm \mu C/ {\rm cm}^{2}$ \\\hline
Important $\bm{k}$-space regions & $\pm K$ (71 $\%$) & $\pm K$ (49 $\%$)/$M_1, M_2, M_3$ (30 $\%$) & $Q, Q'$ (92 $\%$)\\\hline\hline
$P_z (\Phi\neq0)$ &  1.77 $\rm \mu C /{\rm cm}^{2}$  &  0.6 $\rm \mu C /{\rm cm}^{2}$  &  0.034 $\rm \mu C/ {\rm cm}^{2}$  \\\hline
Important $\bm{k}$-space regions & $\pm K$ (71 $\%$) & $\pm K$ (50 $\%$)/$M_1, M_2, M_3$ (31 $\%$) & $Q, Q'$ (98 $\%$)\\\hline\hline
\end{tabular}
\label{table:PzwithRatio}
\end{table*}

\pagebreak

\noindent\textbf{Type I: AB-stacked honeycomb bilayers (SiC)}\\
Parameters used in solving the self-consistent equations in Eq.~\eqref{eq:selfconsistenteqn} are $\epsilon_r=9.66$ for the relative permittivity \cite{eps_sicS}, $d_z=3.5$ {\AA} for the inter-layer distance. All other parameters are the same as those used in obtaining Fig.~\ref{fig:contour_supp}(a). A self-consistent value of $P_z=1.773~\mu$C/cm$^2$ is obtained by solving Eqs.~\eqref{eq:selfconsistenteqn} and presented in Table I of the main text. The momentum-space distribution of polarization is shown in Fig.~2e of the main text where dominant contribution to $P_z$ still arise from $K$ and $K'$ valleys. The band structure of bilayer SiC with a self-consistently obtained $\Phi = -0.073$ eV by solving Eqs.~\eqref{eq:selfconsistenteqn} is shown in Fig.~\ref{fig:fig_supp}.\\

\noindent\textbf{Type II: Rhombohedral (3R) bilayer TMDs (3R-MoS$\bm{_2}$)}\\
Parameters used in solving the self-consistent equations in Eq.~\eqref{eq:selfconsistenteqn} are $\epsilon_r=7$ \cite{Yang2022S} and $d_z=7$ {\AA}, with all other tight-binding parameters being the same as those used in Fig.~\ref{fig:contour_supp}(b).  For this system, we obtain a self-consistent $P_z=-0.6~\mu$C/cm$^2$ and $\Phi = 0.05$ eV  by solving Eqs.~\eqref{eq:selfconsistenteqn}, in agreement with the value reported in experiments~\cite{Yang2022S}. The $\bm{k}$-resolved polarization in the entire Brillouin zone under self-consistent value of $\Phi$ is plotted in Fig.~2(f) of the main text and the band structure of bilayer 3R-MoS$_2$ with $\Phi = 0.05$ eV is shown in Fig.\ref{fig:fig_supp}. \\

\noindent\textbf{Type III: Bilayer T$_{\rm d}$-MX$_2$ (T$_{\rm d}$-WTe$\bm{_2}$)}\\
 Parameters used in solving the self-consistent equations in Eq.~\eqref{eq:selfconsistenteqn} are $\epsilon_r=15$~\cite{eps_tmdS}, and  $d_z=5$ {\AA}, with all other tight-binding parameters kept the same as those used in Fig.~\ref{fig:contour_supp}(c). We obtain a self-consistent value of $P_z=-0.034~\mu$C/cm$^2$ with $\Phi = 1.2$ meV. The $\bm{k}$-resolved eventual polarization $P_z$ in the entire Brillouin zone is presented in Fig.~2(g) of the main text and the band structure is plotted in Fig.~\ref{fig:fig_supp}(c). The polarization as a function of the inversion-symmetric $g_{-}=(g_{pd}-g_{dp})/2$ is calculated self-consistently via Eq.~\eqref{eq:selfconsistenteqn} and presented in Fig.~2(h).\\

Before moving on to the next section, we present a few explanatory notes on the unit conventions for $P_z$ in bilayer systems adopted in the literature:

(i) When “volume” is defined as the conventional 3D volume $V=Ad$ we have the same unit of $P_z$ as in bulk systems - for example, the unit for $P_z$ used in Ref. 13 of the main text is $\mu$C/cm$^2$. We follow this convention throughout this manuscript because it applies in general to any system, regardless of whether it is a bulk material or a 2D material;

(ii) When “volume” is defined as the 2D volume for 2D materials $V\equiv A$, the polarization is given by $P_z = Qd/A$, and the unit would then be given in $C/m$ as adopted in Ref. \cite{WuS}. This definition is only valid for bilayer systems where the separation $D$ between top and bottom surfaces of the system is identical to the inter-layer spacing $d$, and the volume is defined as the total 2D area $A$ of the bilayer system. The polarization is then referred to as the ``2D polarization" $P_{2D}$, \textit{i.e.}, dipole moment over the area of the system. In the specific example of bilayer SiC, the value of $P_{2D}$ taken from Ref.~\cite{WuS} is $P_{2D} = 6.17$ pC/m. 

It is important to note that the conventional 3D definition of polarization $P_z$, \textit{i.e.}, dipole moment over the volume of the system, is related to the 2D polarization by $P_{z} = P_{2D}/d_z$, and this is how we convert the literature value of $P_z$ from Ref.~\cite{WuS} on bilayer SiC to $P_z= 1.76$ $\mu$C/cm$^2$ as presented in Table I of the main text. Similar dimensional analysis for the relation between $P_{2D}$ and $P_z$ applies to other SFE materials.\\

\section*{III. Other contributions to SFE polarization}

In the main text we focus on contributions near the $K$ and $Q$ points within the outermost sets of valence bands in bilayer SFEs. Here we discuss contributions from the other parts of the Brillouin zone, and address briefly the role of deeper lying valence bands that were not included in our lattice Hamiltonians.

\subsection*{1. AB-stacked honeycomb bilayer}

As we explain in the main text, the form of the effective Hamiltonian near $\pm K$ in AB-stacked honeycomb bilayer (Eq.\ 4 of the main text) leads to effective SSH chain in the $N_z \rightarrow \infty$ limit with polar Berry phase given in Eq.\ (3). For $\bm{k}$ away from $\pm K$ points, however, the Hamiltonian no longer takes the form of asymmetrically coupled massive Dirac fermions as in Eq.\ (4). Instead, the dominant term in the microscopic Hamiltonian in Eq.\ \eqref{eq:tb_sic} is given by the intra-layer inter-sublattice hopping $h_{AB}(\bm{k}) = -tf(\bm{k})$, which can be as large as $ \sim -3t = 4.5$ eV for $\bm{k} \simeq 0$ in the neighborhood of the $\rm \Gamma$ point. This term introduces a large $d_{x}\sigma_x$ component in the SSH Hamiltonian Eq.\ (1) with $d_x \gg d_y, d_z$ , which strongly pins the $\bm{d}$-vector along the $x$-axis. The solid angle $\Omega$ subtended in Fig.\ 1(b) and the resultant Berry phase thus become small. This effect is retained in the bilayer limit: as demonstrated in Fig.\ 2(e) of the main text, the $\bm{k}$-resolved polarization is small in the neighborhood of $\rm \Gamma$. 

\subsection*{2. 3R bilayer TMD}

While the effective Hamiltonian of 3R bilayer TMD (Eq.\ \ref{eq:Heff_mos2}) is similar to that of AB-stacked honeycomb bilayer at $\pm K$ (Eq.\ 4 of the main text), the form of the microscopic Hamiltonian in Eq.\ \eqref{eq:H3R} is drastically different from honeycomb bilayers for a general $\bm{k}$. In particular, in the neighborhood of $\rm \Gamma$ we have $f_{\pm}(\bm{k}\simeq 0) \simeq 0$ and $f_0(\bm{k}\simeq 0) \simeq 1$ in $t_{R}(\bm{k})$ (see Eq.\ \ref{eq:interlayer3R}). This indicates that the inter-layer coupling away from $\pm K$ in 3R TMD is dominated by intra-orbital terms while the inter-orbital terms are negligible. $t_{R}(\bm{k})$ becomes an approximately diagonal matrix, \textit{i.e.}, the inter-layer coupling in the band basis also becomes diagonal as the conduction and valence bands are formed by different $d$-orbitals \cite{GuiBinS}. In this case the effective SSH chain has $d_x, d_y \simeq 0$  which results, once again, in small Berry phase. This picture is also verified by the negligible layer polarization near the $\rm \Gamma$-point as shown in Fig.\ 2(f) of the main text.

On the other hand, as we point out in subsection II-2 above, the $\bm{k}$-resolved polariation $P_{z,0}(\bm{k})$ and $P_{z}(\bm{k})$ in Fig.~\ref{fig:contour_supp}(b) and Fig.~\ref{fig:contour_supp}(e) exhibit non-negligible contributions in regions near the time-reversal invariant $M_{1}, M_2, M_3$ points. Here, we show that the effective Hamiltonians near these $M$-points can also be mapped to a two-cell SSH chain, and the bare bilayer polarization $P_{z,0}(\bm{k}\simeq \bm{M})$ is well captured by the polar Berry phase of a similar form in Eq.3 of the main text although the origin of $\Delta_{AB}$ and $g\equiv 2t$ entering the Berry phase is different from the SSH model at the $\pm K$ points.

Before going into detailed model analysis, we first note that thanks to the three-fold $C_{3z}$ symmetry contained in the $C_{3v}$ point group of 3R-MoS$_2$, it suffices to consider one out of the three in-equivalent $M$-points: $M_1 = (0, \frac{2\pi}{\sqrt{3}a}), M_2 = (-\frac{\pi}{a}, -\frac{\pi}{\sqrt{3}a}), M_3 = (\frac{\pi}{a}, -\frac{\pi}{\sqrt{3}a})$ which are related by $C_{3z}$ symmetry: $M_2 = C_{3z}M_1, M_3 = C^2_{3z}M_1$. Note that for a momentum $\bm{k}_1 = \bm{p}_1 + \bm{M}_1$ in the vicinity of $M_1$, $C_{3z}$ requires that $P_{z,0}(\bm{k}_1) = P_{z,0}(C_{3z}\bm{k}_1) = P_{z,0}(C^2_{3z}\bm{k}_1)$, thus the contribution $P_{z,0}(\bm{k}_1)$ from $\bm{k}_1$ is identical to $P_{z,0}(\bm{p}_2 + \bm{M}_2)$ from $\bm{p}_2 + \bm{M}_2$ where $\bm{p}_2 = C_{3z}\bm{p}_1$, and $P_{z,0}(\bm{p}_3 + \bm{M}_3)$ from $\bm{p}_3 + \bm{M}_3$ where $\bm{p}_3 = C^2_{3z}\bm{p}_1$. Suppose the local regions near $M_{i=1,2,3}$ each contributing to a percentage $p_{M_i}$ of the total $P_{z,0}$, $C_{3z}$ guarantees $p_{M_1} = p_{M_2} = p_{M_3}$ and the total contribution from the three $M$-points would be $p_M = \sum_{i=1,2,3} p_{M_i} = 3p_{M_1}$. We thus focus on $\bm{k}$ points near $M_1$ only in the following discussions. 

\begin{figure*}
    \centering
    \includegraphics[width=\linewidth]{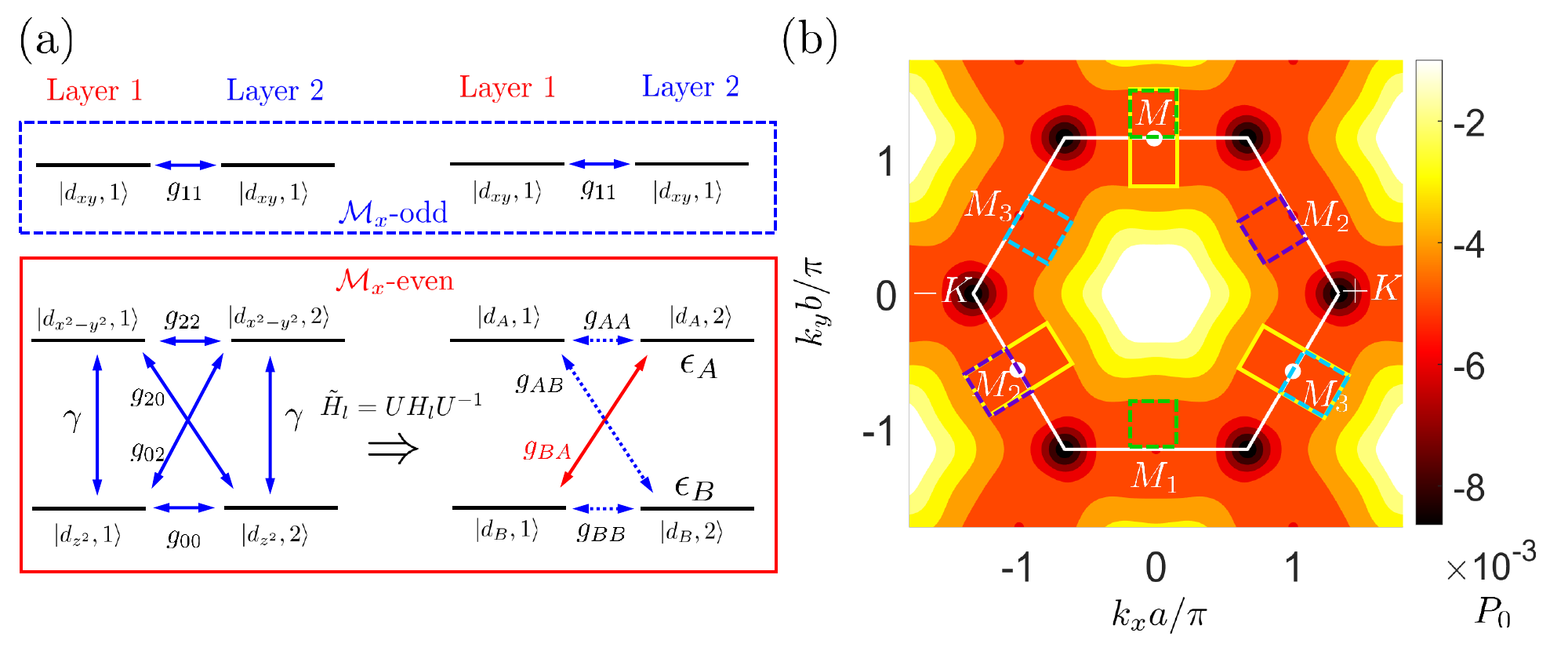}
    \caption{SSH physics at the $M$-points of bilayer 3R-MoS$_2$. (a) Schematic band diagram for states near $M_1$-point. The valence band states result from the mirror($\mathcal{M}_x$)-even sector which forms a two-cell SSH Hamiltonian with broken intra-cell bonding under a proper local unitary transformation $U$ for each monolayer. (b) Enlarged contour plot of $P_{z,0}(\bm{k})$ with regions near time-reversal-invariant $M_1$, $M_2$, $M_3$ points being enclosed by rectangles in which the SSH physics depicted in (a) holds. The two dashed rectangles in green (purple, light blue) denote regions with equivalent $\bm{k}$-points near $M_1$ ($M_2$, $M_3$) as they are related by the reciprocal lattice vector connecting the two equivalent $M_1$ ($M_2$, $M_3$) points. }
    \label{fig:Mpoint_supp}
\end{figure*}

We first consider the special case of $\bm{k} = \bm{M}_1$ which is located along the high-symmetry $\Gamma - M$ line with momentum $\bm{k} = (0, k_y)$ invariant under the vertical mirror plane $\mathcal{M}_{x}: (k_x, k_y) \mapsto (-k_x, k_y)$. Along this  $\mathcal{M}_{x}$-invariant line, the $\mathcal{M}_{x}$-odd $d_{xy}$ orbital is decoupled from the $\mathcal{M}_{x}$-even sectors formed by $d_{z^2}, d_{x^2 - y^2}$ orbitals. In fact, by setting $\bm{k} = \bm{M}_1$ in $H_{3R}(\bm{k})$ in Eq.~\eqref{eq:H3R} we find that the decoupled $\mathcal{M}_{x}$-odd $d_{xy}$ orbitals form the higher-lying conduction bands with energies $E_{d_{xy}}(\bm{M}_1) \gtrsim 2$ eV that are irrelevant for the evaluation of polarization, while the low-energy $\mathcal{M}_{x}$-even sector produces the valence band states that contribute to the polarization as shown in the schematic band diagram in Fig.~\ref{fig:Mpoint_supp}(a). This motivates us to focus on the low-energy Hamiltonian formed by $d_{z^2}, d_{x^2-y^2}$ orbitals which takes the form of a $4\times 4$ matrix per each spin:
\begin{eqnarray}\label{eq:HM}
H_{3R,M} &=& 
\begin{pmatrix}
H_{1, M} & H_{T,M}\\
H^{\dagger}_{T, M} & H_{2, M}
\end{pmatrix},\\\nonumber
H_{1,M} = H_{2, M} &=& 
\begin{pmatrix}
E_{0, M} & \gamma\\
\gamma^{\ast}  & E_{2, M} 
\end{pmatrix}, 
\hspace{5mm} H_{T,M} = 
\begin{pmatrix}
g_{00} & g_{02}\\
g_{20}  & g_{22} 
\end{pmatrix}.
\end{eqnarray}
Note that the matrix elements in Eq.~\eqref{eq:HM} are related to those in $H_{3R}(\bm{k})$ (Eq.~\eqref{eq:H3R}) by: $H_{0,M} = V_0 (\bm{k}=\bm{M}_1)$, $H_{2,M} = V_{22} (\bm{k}=\bm{M}_1)$, $\gamma = V_{2} (\bm{k}=\bm{M}_1)$. The inter-layer terms are $g_{00} = w_0 f_0(\bm{k}=\bm{M}_1)$, $g_{02} = \frac{\gamma_0}{\sqrt{2}} (f_{+}(\bm{k}=\bm{M}_1) + f_{-}(\bm{k}=\bm{M}_1))$, $g_{20} = \frac{\gamma_1}{\sqrt{2}} (f_{+}(\bm{k}=\bm{M}_1) + f_{-}(\bm{k}=\bm{M}_1))$, $g_{22} = w_1 f_0(\bm{k}=\bm{M}_1)$ as in the inter-layer Hamiltonian matrix $t_{R}(\bm{k})$ (Eq.~\eqref{eq:interlayer3R}), which are generally non-vanishing as shown schematically in Fig.~\ref{fig:Mpoint_supp}(a).

To facilitate our model mapping procedures, we perform a local unitary transformation $U$ on each layer which diagonalizes the monolayer Hamiltonians as: 
\begin{eqnarray}
\tilde{H}_{l, M} = U H_{l, M} U^{-1} =  
\begin{pmatrix}
\epsilon_A & 0\\
0 & \epsilon_B
\end{pmatrix}
\hspace{5mm} (l=1,2),
\end{eqnarray}
with $\epsilon_{A} = 2.3$ eV, $\epsilon_{B} = -1.0$ eV obtained by numerically diagonalizing $H_{1,M}, H_{2,M}$ above. $H_{3R,M}$ in Eq.~\eqref{eq:HM} is now transformed as:
\begin{eqnarray}
\tilde{H}_{3R, M} =   
\begin{pmatrix}
\tilde{H}_{l, M} & \tilde{H}_{T, M}\\
\tilde{H}^{\dagger}_{T, M} & \tilde{H}_{l, M}
\end{pmatrix},
\hspace{5mm} \tilde{H}_{T, M} = U H_{T, M} U^{-1} =
\begin{pmatrix}
g_{AA} &  g_{AB}\\
g_{BA} &  g_{BB}
\end{pmatrix}.
\end{eqnarray}
Note that $\ket{d_{A, l=1,2}}$ and $\ket{d_{B, l=1,2}}$ which forms the basis of $\tilde{H}_{3R, M}$ correspond to two orthogonal linear combinations of $d_{z^2}$ and $d_{x^2-y^2}$ orbitals. Interestingly, we numerically find that the transformed inter-layer coupling is dominated by the $g_{BA}$ term, with $|g_{BA}| \approx 0.174 \text{ eV} \gg |g_{AA}|, |g_{AB}|, |g_{BB}| \sim 0.05 \text{ eV}$. This suggests that $\tilde{H}_{3R, M}$ can be approximated by the following effective Hamiltonian:
\begin{eqnarray}\label{eq:SSHM}
\tilde{H}_{\rm{eff}, M} \simeq 
\begin{pmatrix}
\epsilon_{A} & 0 & 0 & 0\\
0 & \epsilon_{B} & g_{BA} & 0 \\
0 & g_{BA} & \epsilon_{A} & 0 \\
0 & 0 & 0 & \epsilon_{B}
\end{pmatrix},
\end{eqnarray}
which is exactly the two-cell SSH chain with broken intra-cell bonding $t+\delta t = 0$ and $g_{BA} = 2t$ as discussed in the main text for the $\pm K$ points in AB honeycomb bilayer and 3R-MoS$_2$. To further justify that the approximation in $\tilde{H}_{\rm{eff}, M}$ in Eq.~\eqref{eq:SSHM} above has well captured the polarization physics, we note that following exactly the same analysis presented in Eq.~\eqref{eq:PKBerryPhase} in subsection II.2A, the polarization for $\tilde{H}_{\rm{eff}, M}$ above is given by $P_{z,0}(\bm{M}_1) = -\frac{\gamma_M}{2\pi} P_0$, with 
\begin{equation}
\gamma_M = \pi (1 - \frac{\Delta_{AB}}{\sqrt{\Delta^2_{AB} + g^2_{BA}}}),
\end{equation}
which takes exactly the same form of Eq.3 of the main text. With $\epsilon_{A} = 2.3$ eV, $\epsilon_{B} = -1.0$ eV, $\Delta_{AB} = \frac{\epsilon_{A}-\epsilon_{B}}{2}$, $|g_{BA}| = 0.174 \text{ eV}$, we obtain $\gamma_{M} = 0.0054 \pi$ for each spin. Thus, the total $P_{z,0}(\bm{M}_1)$ obtained from $\tilde{H}_{\rm{eff}, M}$ with both spins included would be $P_{z,0}(\bm{M}_1) = -2 \frac{\gamma_{M}}{2\pi} P_0 = -0.0054 P_0$, which has an excellent agreement with the value $P_{z,0}(\bm{M}_1) = -0.005 P_0$ we obtained numerically from the realistic lattice model (Eq.~\eqref{eq:H3R}) at $\bm{k} = \bm{M}_1$ using the formula for $P_{z,0}(\bm{k})$ in Eq.~\eqref{eq:sumPk}.

For a general $\bm{k} \neq \bm{M}_1$, we confirm numerically that within the region $ -0.17 {\AA}^{-1} < k_x < 0.17 {\AA}^{-1}, 0.514 {\AA}^{-1} < k_y < 1.754 {\AA}^{-1}$ (rectangular area enclosed by solid yellow lines surrounding $M_1$ point in Fig.~\ref{fig:Mpoint_supp}(b)), the matrix elements $V_{1}$, $V_{12}$ in $H_{3R}(\bm{k})$ (Eq.~\eqref{eq:H3R}) which mix the $d_{xy}$ orbitals with $d_{z^2}, d_{x^2-y^2}$ orbitals generally satisfy $|V_{1}| \sim |V_{12}| \ll \gamma \equiv |V_{2}| $. Thus, the $d_{xy}$ orbital remains largely decoupled from the $d_z^2, d_{x^2 - y^2}$ sector throughout this rectangular region , and the schematic band diagram depicted in Fig.~\ref{fig:Mpoint_supp}(a) as well as the form of the effective Hamiltonian in Eq.~\eqref{eq:SSHM} still holds approximately for a general $\bm{k} \simeq \bm{M}_1$. This explains why $P_{z,0}(\bm{k})$ remains almost uniform within the rectangular region enclosed by the yellow lines around $M_1$. We verify numerically that $0.85<|\gamma_{\bm{k}}|/|\gamma_M|<1.02$ throughout the rectangular region around $M_1$, where $\gamma_{\bm{k}} \equiv -\pi P_{z,0}(\bm{k})/P_0$ with $P_{z,0}(\bm{k})$ obtained numerically for a general $\bm{k}$ around $M_1$. The analyses above confirm the origin of $P_{z,0}(\bm{k})$ near $M$-points to be well captured by $\gamma_M$ derived from the SSH model in Eq.~\eqref{eq:SSHM}.

The contribution from regions near all three $M_1, M_2, M_3$ points is calculated by summing over contributions within the three rectangular regions enclosed by yellow solid lines in Fig.~\ref{fig:Mpoint_supp}(b) in which the SSH physics discussed above holds. We note that the three rectangles in yellow solid lines in Fig.~\ref{fig:Mpoint_supp}(b) have faithfully captured all contributions near $M_1, M_2, M_3$ within the first Brillouin zone (hexagonal area enclosed by the white solid line) - the set of $\bm{k}$-points within the green dashed rectangle above $M_1 = (0, -\frac{2\pi}{\sqrt{3}a})$ are equivalent to the set of $\bm{k}$-points within the green dashed rectangle above $M_1 = (0, +\frac{2\pi}{\sqrt{3}a})$ as they are related by the reciprocal lattice vector $\bm{G}_1 = (0, \frac{4\pi}{\sqrt{3}a})$ connecting the two equivalent $M_1$ points. Same analysis applies to the two rectangles in purple dashed lines near the two equivalent $M_2$ points as well as the two in light blue dashed lines near the two $M_3$ points. The total contribution from the neighborhood of all three $M$-points is found numerically to be $p_M = 3 p_{M_1} \approx 30\%$ of the total $P_{z,0}$ (see Table~\ref{table:PzwithRatio}).

\subsection*{3. T$_{\rm d}$ bilayer TMD}

In the derivation of the effective coupled massive Dirac fermions in Eq.\ \eqref{eq:CoupledMassiveDirac} we focus on $\bm{k} \simeq Q, Q'$ where the topological band crossing occurs between $p$,$d$-orbitals, \textit{i.e.}, $E_p(\bm{k} = \bm{Q}) = E_d(\bm{k} = \bm{Q})$ in Eq.\ \eqref{eq:monolayerWTe2}. For $\bm{k}$ away from $Q, Q'$, however, the energy difference $\Delta(\bm{k}) = E_p(\bm{k}) - E_d(\bm{k})$ becomes significant, which can be of order 2 eV near $\rm \Gamma$ and $>3$ eV near $\rm X$ according to Fig.~\ref{fig:fig_supp} and thus dominate over all other energy scales given by SOC ($\sim 0.1$ eV) and inter-layer coupling ($\sim 0.01 - 0.1$ eV). Note that $\Delta(\bm{k})$ enters the $\mathbf{k}\cdot\mathbf{p}$ Hamiltonian in Eq.\ref{eq:Hkdotp} as $\frac{\Delta(\bm{k})}{2} s_z \sigma_0$, which upon the basis transformation introduced in Eq.\ \eqref{eq:basistrans} becomes $\frac{\Delta(\bm{k})}{2} s_x \sigma_0$. This term introduces a large $d_x \sigma_x$ term in the spinful SSH chain in Eq.\ \eqref{eq:spinfulSSH} and strongly pins the $\bm{d}_{\xi\sigma}$ vector to the $x$-axis of the Bloch sphere. The resultant Berry phase is then negligible due to the small solid angle $\Omega$ subtended by $\bm{d}_{\xi\sigma}$. As we confirm in Fig.\ 2(g), the layer polarization is negligible across the Brillouin zone except near the band crossing points at $Q, Q'$.

\subsection*{4. Role of deeper lying valence bands}

Finally, we address briefly the role of deeper lying valence bands which are not included in the lattice Hamiltonians in Eq.~\eqref{eq:tb_sic}, Eq.~\eqref{eq:H3R} and Eq.~\eqref{eq:tb_tmd}. In principle, the polarization of a material should be given by summing over Berry phases from all filled valence bands. On the other hand, we note that our microscopic model calculations which take into account the outermost sets of valence bands have produced accurate values of $P_z$ in excellent agreement with literature values as shown in Table I of the main text. We believe this justifies our lattice formalism with outermost sets of valence bands as a reasonably good description polarization physics in bilayer SFEs discussed in this manuscript, and inclusion of deeper lying valence bands is not absolutely essential.

While a thorough investigation into the role of deeper lying valence electrons requires detailed \textit{ab initio} calculations beyond the scope of the current work, we discuss possible physical reasons behind the relatively minor contributions from deeper lying valence electrons. The valence electrons in AB-stacked honeycomb bilayers such as hBN and SiC are known to consist of the three $p_x, p_y, p_z$-orbitals: the $p_x, p_y$-orbitals go through the famous $sp2$ hybridization to form strong covalent bonds with neighboring atoms which stabilizes the overall 2D planar structure~\cite{ChabiS}; on the other hand, the $p_z$ orbitals are not involved in the strong covalent bonding - they form the $\pi$ bands that are well captured by the lattice Hamiltonian we considered in Eq.~\eqref{eq:tb_sic}. Upon van der Waals stacking of two monolayers, electron clouds in the deep lying valence bands from strong $sp2$ hybridization are expected to be largely insensitive to the much weaker inter-layer van der Waals interactions, thus not likely to contribute significantly to the stacking polarization. In contrast, being free from strong $sp2$ hybridization, the electron clouds associated with the $p_z$-orbitals can deform more easily under van der Waals stacking and gives rise to the stacking polarization.

The deeper lying valence bands in bilayer 3R-TMDs also arise from strong $pd$ hybridization between $p$ orbitals from chalcogen atoms X=S, Se and a subset of $d$ orbitals from transition-metal M = Mo, W atoms (see orbital composition of deep lying valence bands in Ref.~\cite{GuiBinS}), which forms the strong covalent bonds that stabilize the overall prismatic atomic coordination within each monolayer~\cite{LongoS}. Following the same line of reasoning above, the strongly $pd$ hybridized valence electrons are not expected to play a significant role in the stacking polarization. On the other hand, the topmost valence bands are dominated by the M-$d$ orbitals with minor X-$p$ orbital composition~\cite{GuiBinS}, indicating weak $pd$ hybridization within these bands  - these weakly bonded $d$-bands, as captured by our lattice model, are more susceptible to deformations under van der Waals forces, thus contributing the most to the stacking polarization.


\begin{thebibliography}{99}

\bibitem{Garcia} Vincent Garcia and Manuel Bibes, \href{https://www.nature.com/articles/ncomms5289}{Nat. Commun. {\bf 5}, 4289 (2014)}.

\bibitem{Datta} Asif Islam Khan, Ali Keshavarzi and Suman Datta, \href{https://www.nature.com/articles/s41928-020-00492-7}{Nat. Electronics {\bf 3}, 588–597 (2020)}.

\bibitem{Wu} M. Wu and J. Li, \href{https://www.pnas.org/doi/10.1073/pnas.2115703118}{Proc. Natl. Acad. Sci. USA {\bf 118}, (50) e2115703118 (2021)}.

\bibitem{Li} L. Li and M. Wu, \href{https://pubs.acs.org/doi/10.1021/acsnano.7b02756}{ACS Nano {\bf 11}, 6382 (2017)}.

\bibitem{Fei} Z. Fei, W. Zhao, T. A. Palomaki, B. Sun, M. K. Miller, Z. Zhao, J. Yan, X. Xu, and D. H. Cobden, \href{https://www.nature.com/articles/s41586-018-0336-3}{Nature {\bf 560}, 336 (2018)}.

\bibitem{DelaBarrera} S. C. de la Barrera, Q. Cao, Y. Gao, Y. Gao, V. S.
Bheemarasetty, J. Yan, D. G. Mandrus, W. Zhu, D. Xiao, and B. M. Hunt, \href{https://www.nature.com/articles/s41467-021-25587-3}{Nat. Commun. {\bf 12}, 5298 (2021)}.

\bibitem{Yasuda} Kenji Yasuda, Xirui Wang, Kenji Watanabe, Takashi Taniguchi, Pablo Jarillo-Herrero, \href{https://www.science.org/doi/full/10.1126/science.abd3230?rss=1}{Science 372, 6549 (2021)}.

\bibitem{Stern} Maayan Vizner Stern, Yuval Waschitz, Wei Cao, Iftach Nevo, Kenji Watanabe, Takashi Taniguchi, Eran Sela, Michael Urbakh, Oded Hod, Moshe Ben Shalom, \href{https://www.science.org/doi/10.1126/science.abe8177}{Science {\bf 372}, 6549 (2021)}.

\bibitem{Jindal} A. Jindal, A. Saha, Z. Li, T. Taniguchi, K. Watanabe, J. C. Hone, T. Birol, R. M. Fernandes, C. R. Dean, A. N. Pasupathy, and D. A. Rhodes, \href{https://www.nature.com/articles/s41586-022-05521-3}{Nature {\bf 613}, 48 (2023)}.

\bibitem{Xirui} Xirui Wang, Kenji Yasuda, Yang Zhang, Song Liu, Kenji Watanabe, Takashi Taniguchi, James Hone, Liang Fu and Pablo Jarillo-Herrero, \href{https://www.nature.com/articles/s41565-021-01059-z}{Nat. Nanotech. {\bf 17}, 367–371 (2022)}.

\bibitem{Yang} D. Yang, J. Wu, B. T. Zhou, J. Liang, T. Ideue, T. Siu, K. M. Awan, K. Watanabe, T. Taniguchi, Y. Iwasa, M. Franz, and Z. Ye, \href{https://www.nature.com/articles/s41566-022-01008-9}{Nat. Photonics {\bf 16}, 469 (2022)}.

\bibitem{Jing} Jing Liang, Dongyang Yang, Jingda Wu, Jerry I. Dadap, Kenji Watanabe, Takashi Taniguchi, and Ziliang Ye, \href{https://journals.aps.org/prx/abstract/10.1103/PhysRevX.12.041005}{Phys. Rev. X {\bf 12}, 041005 (2022)}.

\bibitem{Zheng} Z. Zheng, Q. Ma, Z. Bi, S. de la Barrera, M. H. Liu, N. Mao, Y. Zhang, N. Kiper, K. Watanabe, T. Taniguchi, J. Kong, W. A. Tisdale, R. Ashoori, N. Gedik, L. Fu, S. Y. Xu, and P. Jarillo-Herrero, \href{https://www.nature.com/articles/s41586-020-2970-9}{Nature {\bf 588}, 71 (2020)}.

\bibitem{Niu} R. Niu, Z. Li, X. Han, Z. Qu, D. Ding, Z. Wang, Q. Liu, T. Liu, C. Han, K. Watanabe, T. Taniguchi, M. Wu, Q. Ren, X. Wang, J. Hong, J. Mao, Z. Han, K. Liu, Z. Gan, and J. Lu, \href{https://www.nature.com/articles/s41467-022-34104-z}{Nat. Commun. {\bf 13}, 6241 (2022)}.

\bibitem{Pacchioni} Giulia Pacchioni, \href{https://www.nature.com/articles/s41578-022-00525-x}{Nat. Rev. Mater. {\bf 8}, 8 (2023)}.

\bibitem{Ji} J. Ji, C. Xu, and H. J. Xiang, \href{https://journals.aps.org/prl/abstract/10.1103/PhysRevLett.130.146801}{Phys. Rev. Lett. {\bf 130}, 146801 (2023)}.

\bibitem{Vanderbuilt} R. D. King-Smith and David Vanderbilt, \href{https://journals.aps.org/prb/abstract/10.1103/PhysRevB.47.1651}{Phys. Rev. B {\bf 47}, 1651 (R) (1993)}.

\bibitem{Resta} Raffaele Resta, \href{https://journals.aps.org/rmp/abstract/10.1103/RevModPhys.66.899}{Rev. Mod. Phys. {\bf 66}, 899 (1994)}

\bibitem{Resta2} R Resta, D Vanderbilt, \href{https://link.springer.com/chapter/10.1007/978-3-540-34591-6_2}{Physics of Ferroelectrics (2007)}.

\bibitem{Xiao} D. Xiao, M. C. Chang, and Q. Niu, \href{https://journals.aps.org/rmp/abstract/10.1103/RevModPhys.82.1959}{Rev. Mod. Phys. {\bf 82}, 1959 (2010)}.

\bibitem{Spaldin} N. A. Spaldin,  \href{https://www.sciencedirect.com/science/article/abs/pii/S0022459612003234?via%3Dihub}{Journal of Solid State Chemistry {\bf 195}, 2 (2012)}.

\bibitem{Fregoso} S. R. Panday and B. M. Fregoso, \href{https://iopscience.iop.org/article/10.1088/1361-648X/aa8bfc}{J. Phys.: Condens. Matter {\bf 29} 43LT01, (2017)}.

\bibitem{CZheng} C. Zheng et al., \href{https://www.science.org/doi/10.1126/sciadv.aar7720}{Sci. Adv. {\bf 4}, 7 (2018)}.

\bibitem{Higashi} N. Higashitarumizu et al., \href{https://www.nature.com/articles/s41467-020-16291-9}{Nat. Commun. {\bf 11}, 2428 (2020)}.

\bibitem{RestaPRL} R. Resta, \href{https://journals.aps.org/prl/abstract/10.1103/PhysRevLett.80.1800}{Phys. Rev. Lett. {\bf 80}, 1800 (1998)}.

\bibitem{Bennett1} D. Bennett and B. Remez, \href{https://www.nature.com/articles/s41699-021-00281-6}{npj 2D Mat. Appl. {\bf 6}, 7 (2022)}.

\bibitem{Bennett2} D. Bennett, W. J. Jankowski, G. Chaudhary, E. Kaxiras, R.-J. Slager, \href{https://journals.aps.org/prresearch/abstract/10.1103/PhysRevResearch.5.033216}{Phys. Rev. Research {\bf 5}, 033216 (2023)}.

\bibitem{SSH} W. P. Su, J. R. Schrieffer, and A. J. Heeger, \href{https://journals.aps.org/prl/abstract/10.1103/PhysRevLett.42.1698}{Phys. Rev. Lett. {\bf 42}, 1698 (1979)}.

\bibitem{RiceMele} M. J. Rice and E. J. Mele, \href{https://journals.aps.org/prl/abstract/10.1103/PhysRevLett.49.1455}{Phys. Rev. Lett. {\bf 49}, 1455 (1982)}.

\bibitem{Maschmeyer} Jyah Strachan, Anthony F. Masters, and Thomas Maschmeyer, \href{https://pubs.acs.org/doi/10.1021/acsaem.1c00638}{ACS Appl. Energy Mater. {\bf 4}, 8, 7405–7418 (2021)}.

\bibitem{GuiBin} Gui-Bin Liu, Wen-Yu Shan, Yugui Yao, Wang Yao, and Di Xiao, \href{https://journals.aps.org/prb/abstract/10.1103/PhysRevB.88.085433}{Phys. Rev. B {\bf 88}, 085433 (2013)}.

\bibitem{Yao} Di Xiao, Gui-Bin Liu, Wanxiang Feng, Xiaodong Xu, and Wang Yao, \href{https://journals.aps.org/prl/abstract/10.1103/PhysRevLett.108.196802}{Phys. Rev. Lett. {\bf 108}, 196802 (2012)}.

\bibitem{Kormanyos} Andor Korm\'{a}nyos, Viktor Z\'{o}lyomi, Vladimir I. Fal'ko, and Guido Burkard, \href{https://doi.org/10.1103/PhysRevB.98.035408}{Phys. Rev. B {\bf 98}, 035408 (2018)}.

\bibitem{Yao2} Yong Wang, Zhan Wang, Wang Yao, Gui-Bin Liu, and Hongyi Yu, \href{https://journals.aps.org/prb/abstract/10.1103/PhysRevB.95.115429}{Phys. Rev. B {\bf 95}, 115429 (2017)}.

\bibitem{SM} See Supplemental Material for details on (I) microscopic models of SFE materials and effective models at $K$ and $Q$; (II) Relation between Berry phase in periodic SSH chain and origin of $P_z$ in bilayer SFE, which includes: 1. definition of bare polarization $P_{z,0}$; 2. relation between $P_{z,0}$ and Berry phase in periodic SSH chain and dominance of contributions from $K$, $Q$ points; 3. details of self-consistent equation for $P_z$ (Eq.\ \ref{eq:SelfConEqn}); (III) discussions on Berry phases away from $K$ and $Q$ points and deeper lying valence bands.

\bibitem{Car} Lukas Muechler, A. Alexandradinata, Titus Neupert, and Roberto Car, \href{https://journals.aps.org/prx/abstract/10.1103/PhysRevX.6.041069}{Phys. Rev. X {\bf 6}, 041069 (2016)}.

\bibitem{Qiong} Qiong Ma, Su-Yang Xu, Huitao Shen, David Macneill, Valla Fatemi, Andres M. Mier Valdivia, Sanfeng Wu, Tay-Rong Chang, Zongzheng Du, Chuang-Han Hsu, Quinn D. Gibson, Shiang Fang, Efthimios Kaxiras, Kenji Watanabe, Takashi Taniguchi, Robert J. Cava, Hai-Zhou Lu, Hsin Lin, Liang Fu, Nuh Gedik, Pablo Jarillo-Herrero, \href{https://www.nature.com/articles/s41586-018-0807-6}{Nature {\bf 565}, 337–342 (2019)}.

\bibitem{Sanfeng} Sanfeng Wu, Valla Fatemi, Quinn D. Gibson, Kenji Watanabe, Takashi Taniguchi, Robert J. Cava, Pablo Jarillo-Herrero, \href{https://www.science.org/doi/10.1126/science.aan6003}{Science {\bf 359} (6371), 76-79 (2018)}.

\bibitem{Du} Z. Z. Du, C. M. Wang, Hai-Zhou Lu, and X. C. Xie, \href{https://journals.aps.org/prl/abstract/10.1103/PhysRevLett.121.266601}{Phys. Rev. Lett. {\bf 121}, 266601 (2018)}.

\bibitem{Junwei} Xiaofeng Qian, Junwei Liu, Liang Fu, Ju Li, \href{https://www.science.org/doi/10.1126/science.1256815}{Science {\bf 346}, 1344-1347 (2014)}.

\bibitem{Benjamin} Ying-Ming Xie, Benjamin T. Zhou, and K. T. Law, \href{https://journals.aps.org/prl/abstract/10.1103/PhysRevLett.125.107001}{Phys. Rev. Lett. {\bf 125}, 107001 (2020)}.

\bibitem{McCann} Edward McCann, Mikito Koshino, \href{https://iopscience.iop.org/article/10.1088/0034-4885/76/5/056503}{Rep. Prog. Phys. {\bf 76}, 056503 (2013)}.

\bibitem{Burkard} A. Kormanyos, V. Zolyomi, V. I. Fal'ko, and G. Burkard, \href{https://journals.aps.org/prb/abstract/10.1103/PhysRevB.98.035408}{Phys. Rev. B {\bf 98}, 035408 (2018)}.

\bibitem{Rhodes} Apoorv Jindal, Amartyajyoti Saha, Zizhong Li, Takashi Taniguchi, Kenji Watanabe, James C. Hone, Turan Birol, Rafael M. Fernandes, Cory R. Dean, Abhay N. Pasupathy \& Daniel A. Rhodes, \href{https://www.nature.com/articles/s41586-022-05521-3}{Nature {\bf 613}, 48–52 (2023)}.

\bibitem{Crepel} V. Cr\'{e}pel and L. Fu, \href{https://www.pnas.org/doi/10.1073/pnas.2117735119}{Proc. Natl. Acad. Sci. USA {\bf 119}, e2117735119 (2022)}.

\bibitem{Rappe} Liang Z. Tan, Fan Zheng, Steve M Young, Fenggong Wang, Shi Liu and Andrew M Rappe, \href{https://www.nature.com/articles/npjcompumats201626}{npj Computational Materials {\bf 2}, 16026 (2016)}.

\bibitem{Moore} Benjamin M. Fregoso, Takahiro Morimoto, and Joel E. Moore, \href{https://journals.aps.org/prb/abstract/10.1103/PhysRevB.96.075421#:~:text=Electric%20polarization%20has%20a%20quantized,each%20point%20in%20momentum%20space.}{Phys. Rev. B {\bf 96}, 075421 (2017)}.

\bibitem{Cook} Ashley M. Cook, Benjamin M. Fregoso, Fernando de Juan, Sinisa Coh and Joel E. Moore, \href{https://www.nature.com/articles/ncomms14176}{Nat. Commun. {\bf 8}, 14176 (2017)}.

\bibitem{Hongyi} H. Yu, Z. Zhou and W. Yao, \href{https://link.springer.com/article/10.1007/s11433-023-2163-3}{Sci. China Phys. Mech. Astron. {\bf 66}, 107711 (2023)}.
    
\end{thebibliography}

\begin{thebibliography}{99}

\bibitem{sic_dftS}{Lin, X., Lin, S., Xu, Y., \& Chen, H. (2015). Electronic structures of multilayer two-dimensional silicon carbide with oriented misalignment. \href{https://doi.org/10.1039/c5tc01679g}{Journal of Materials Chemistry C, 3(35), 9057–9062.}}

\bibitem{SSHS} W. P. Su, J. R. Schrieffer, and A. J. Heeger, \href{https://doi.org/10.1103/PhysRevB.22.2099}{Phys. Rev. B {\bf 22}, 2099 (1980)}.

\bibitem{GuiBinS} Gui-Bin Liu, Wen-Yu Shan, Yugui Yao, Wang Yao, and Di Xiao, \href{https://journals.aps.org/prb/abstract/10.1103/PhysRevB.88.085433}{Phys. Rev. B {\bf 88}, 085433 (2013)}.

\bibitem{Yang2022S} Yang, D., Wu, J., Zhou, B. T., Liang, J., Ideue, T., Siu, T., Awan, K. M., Watanabe, K., Taniguchi, T., Iwasa, Y., Franz, M., \& Ye, Z.  \href{https://doi.org/10.1038/s41566-022-01008-9}{Nature Photonics, 16(6), 469–474 (2022)}. 

\bibitem{HongyiS} Yong Wang, Zhan Wang, Wang Yao, Gui-Bin Liu, and Hongyi Yu, \href{https://doi.org/10.1103/PhysRevB.95.115429}{Phys. Rev. B {\bf 95}, 115429 (2017)}.

\bibitem{DiXiaoS} Di Xiao, Gui-Bin Liu, Wanxiang Feng, Xiaodong Xu, and Wang Yao, \href{https://journals.aps.org/prl/abstract/10.1103/PhysRevLett.108.196802}{Phys. Rev. Lett. {\bf 108}, 196802 (2012)}.

\bibitem{CarS} Lukas Muechler, A. Alexandradinata, Titus Neupert, and Roberto Car, \href{https://journals.aps.org/prx/abstract/10.1103/PhysRevX.6.041069}{Phys. Rev. X {\bf 6}, 041069 (2016)}.

\bibitem{JunweiS} X. Qian, J. Liu, L. Fu, and J. Li, \href{https://www.science.org/doi/10.1126/science.1256815}{Science {\bf 346}, 1344 (2014)}.

\bibitem{BenjaminS} Ying-Ming Xie, Benjamin T. Zhou, and K. T. Law, \href{https://journals.aps.org/prl/abstract/10.1103/PhysRevLett.125.107001}{Phys. Rev. Lett. {\bf 125}, 107001 (2020)}.

\bibitem{QiongS} Qiong Ma, Su-Yang Xu, Huitao Shen, David Macneill, Valla Fatemi, Andres M. Mier Valdivia, Sanfeng Wu, Tay-Rong Chang, Zongzheng Du, Chuang-Han Hsu, Quinn D. Gibson, Shiang Fang, Efthimios Kaxiras, Kenji Watanabe, Takashi Taniguchi, Robert J. Cava, Hai-Zhou Lu, Hsin Lin, Liang Fu, Nuh Gedik, Pablo Jarillo-Herrero, \href{https://www.nature.com/articles/s41586-018-0807-6}{Nature {\bf 565}, 337–342 (2019)}.

\bibitem{DuS} Z. Z. Du, C. M. Wang, Hai-Zhou Lu, and X. C. Xie, \href{https://journals.aps.org/prl/abstract/10.1103/PhysRevLett.121.266601}{Phys. Rev. Lett. {\bf 121}, 266601 (2018)}.

\bibitem{VanderbuiltS} R. D. King-Smith and David Vanderbilt, \href{https://doi.org/10.1103/PhysRevB.47.1651}{Phys. Rev. B {\bf 47}, 1651(R) (1993)}.

\bibitem{RestaS} R Resta, D Vanderbilt, \href{https://link.springer.com/chapter/10.1007/978-3-540-34591-6_2}{Physics of Ferroelectrics, 31-68 (2007)}.

\bibitem{ZhengS} Z. Zheng, Q. Ma, Z. Bi, S. de la Barrera, M. H. Liu, N. Mao, Y. Zhang, N. Kiper, K. Watanabe, T. Taniguchi, J. Kong, W. A. Tisdale, R. Ashoori, N. Gedik, L. Fu, S. Y. Xu, and P. Jarillo-Herrero, \href{https://www.nature.com/articles/s41586-020-2970-9}{Nature {\bf 588}, 71 (2020)}.

\bibitem{HongyiS} H. Yu, Z. Zhou and W. Yao, \href{https://link.springer.com/article/10.1007/s11433-023-2163-3}{Sci. China Phys. Mech. Astron. {\bf 66}, 107711 (2023)}.

\bibitem{eps_sicS} Chabi, S., \& Kadel, K., Two-dimensional silicon carbide: Emerging direct band gap semiconductor. \href{https://doi.org/10.3390/nano10112226}{Nanomaterials, 10(11), 1–20, (2020)}.

\bibitem{WuS} M. Wu and J. Li, \href{https://www.pnas.org/doi/10.1073/pnas.2115703118}{Proc. Natl. Acad. Sci. USA {\bf 118}, (50) e2115703118 (2021)}.

\bibitem{eps_tmdS}{Domozhirova, A. N., Makhnev, A. A., Shreder, E. I., Naumov, S. V., Lukoyanov, A. V., Chistyakov, V. V., Huang, J. C. A., Semiannikova, A. A., Korenistov, P. S., \& Marchenkov, V. V. (2019). Electronic properties of WTe2 and MoTe2 single crystals. \href{https://doi.org/10.1088/1742-6596/1389/1/012149}{Journal of Physics: Conference Series, 1389(1).}}

\bibitem{ChabiS} Sakineh Chabi, Zeynel Guler, Adrian J. Brearley, Angelica D. Benavidez and Ting Shan Luk, \href{https://www.mdpi.com/2079-4991/11/7/1799}{Nanomaterials 2021, 11(7), 1799 (2021)}.

\bibitem{LongoS} Roberto C. Longo et al., \href{https://iopscience.iop.org/article/10.1088/2053-1583/aa636c/meta}{2D Mater. 4, 025050 (2017)}.

\end{thebibliography}
\end{document}